\documentclass[12pt,aps,prd,preprint,tightenlines,superscriptaddress,
   showpacs,nofootinbib,floatfix]{revtex4}
\newcommand{\PRE}[1]{{#1}} 

\usepackage{bm} 
\usepackage{epsfig}
\usepackage{graphicx}
\usepackage{subfigure}

\newcommand{\ifb}{\text{fb}^{-1}}

\newcommand{\gev}{\text{GeV}}
\newcommand{\tev}{\text{TeV}}

\newcommand{\eg}{{\em e.g.}}
\newcommand{\ie}{{\em i.e.}}

\newcommand{\eqref}[1]{Eq.~(\ref{#1})}

\newcommand{\secref}[1]{Sec.~\ref{sec:#1}}

\newcommand{\appref}[1]{Appendix~\ref{sec:#1}}
\newcommand{\figref}[1]{Fig.~\ref{fig:#1}}
\newcommand{\figsref}[2]{Figs.~\ref{fig:#1} and \ref{fig:#2}}
\newcommand{\Figref}[1]{Figure~\ref{fig:#1}}
\newcommand{\tableref}[1]{Table~\ref{table:#1}}
\newcommand{\tablesref}[2]{Tables~\ref{table:#1} and \ref{table:#2}}

\newcommand{\slepton}{\tilde{\ell}}

\newcommand{\mmess}{M_{\text{m}}}

\newcommand{\nmess}{N_5}

\newcommand{\no}{\nonumber}

\newcommand{\disabled}[1]{{}}

\begin{document}

\preprint{UCI-TR-2009-11}

\title{ \PRE{\vspace*{1.5in}} Measuring Slepton Masses and Mixings 
at the LHC \PRE{\vspace*{0.3in}} }

\author{Jonathan L.~Feng}
\affiliation{Department of Physics and Astronomy, University of
California, Irvine, California 92697, USA}

\author{Sky T.~French}
\affiliation{Cavendish Laboratory, J.~J.~Thomson Avenue, Cambridge,
CB3 0HE, UK}

\author{Iftah Galon}
\affiliation{Physics Department, Technion-Israel Institute of
Technology, Haifa 32000, Israel}

\author{Christopher G.~Lester}
\affiliation{Cavendish Laboratory, J.~J.~Thomson Avenue, Cambridge,
CB3 0HE, UK}

\author{Yosef Nir}
\affiliation{Department of Particle Physics, Weizmann Institute of
Science, Rehovot 76100, Israel \PRE{\vspace*{.5in}} }

\author{Yael Shadmi}
\affiliation{Department of Physics and Astronomy, University of
California, Irvine, California 92697, USA}
\affiliation{Physics Department, Technion-Israel Institute of
Technology, Haifa 32000, Israel}

\author{David Sanford}
\affiliation{Department of Physics and Astronomy, University of
California, Irvine, California 92697, USA}

\author{Felix Yu\PRE{\vspace*{.5in}} }
\affiliation{Department of Physics and Astronomy, University of
California, Irvine, California 92697, USA}

\date{November 2009}

\begin{abstract}
\PRE{\vspace*{.3in}} Flavor physics may help us understand theories
beyond the standard model. In the context of supersymmetry, if we can
measure the masses and mixings of sleptons and squarks, we may learn
something about supersymmetry and supersymmetry breaking. Here we
consider a hybrid gauge-gravity supersymmetric model in which the
observed masses and mixings of the standard model leptons are
explained by a $\text{U(1)} \times \text{U(1)}$ flavor symmetry.  In
the supersymmetric sector, the charged sleptons have reasonably large
flavor mixings, and the lightest is metastable.  As a result,
supersymmetric events are characterized not by missing energy, but by
heavy metastable charged particles.  Many supersymmetric events are
therefore fully reconstructible, and we can reconstruct most of the
charged sleptons by working up the long supersymmetric decay
chains. We obtain promising results for both masses and mixings, and
conclude that, given a favorable model, precise measurements at the
LHC may help shed light not only on new physics, but also on the
standard model flavor parameters.
\end{abstract}


\maketitle

\section{Introduction}
\label{sec:intro}

The study of flavor physics is interesting both because we do not
understand why there is smallness and hierarchy in the standard model
(SM) flavor parameters, and because flavor physics holds a key to
understanding theories beyond the SM.  If supersymmetry is discovered
at the Large Hadron Collider (LHC), then studying its flavor
properties --- the masses and mixings of sleptons and squarks --- may
indeed take us a long way toward understanding the SM flavor
parameters, and will also shed light on the underlying structure of
supersymmetry and supersymmetry breaking.  In this work, we take some
new steps toward a quantitative analysis of the actual prospects for
supersymmetric flavor measurements at the LHC, following earlier
studies of lepton flavor studies at the LHC~\cite{ArkaniHamed:1996au,%
ArkaniHamed:1997km,Agashe:1999bm,Hisano:2002iy,Goto:2004cpa,%
Kitano:2008en,Allanach:2008ib,Kaneko:2008re,DeSimone:2009ws}.  Our
goal is to answer the following questions:
\begin{itemize}
\item How many sleptons might the LHC be able to identify?
\item How precisely can the masses of these sleptons be measured?
\item Can the flavor decomposition of these sleptons be determined?
\end{itemize}
The answers to these questions may ultimately allow us to identify the
correct theory of lepton flavor, which has eluded us for decades
despite plentiful experimental data.  Still, even these first
questions are ambitious, and for this reason our objective here is not
to undertake a comprehensive study of all supersymmetric models, but
rather to work toward an existence proof that, in some well-motivated
cases, significant progress is possible.  To do this, we will choose a
favorable model to analyze.

In Ref.~\cite{Feng:2007ke}, hybrid supersymmetric models were
constructed in which sfermion masses receive both flavor-conserving
gauge-mediated contributions and flavor-violating gravity-mediated
contributions governed by a $\text{U(1)} \times \text{U(1)}$
horizontal symmetry.  These models satisfy all low-energy constraints
obtained from flavor factories, and also explain all charged lepton
and neutrino masses and mixings in terms of a few charge assignments
under the flavor-symmetry.  At the same time, the gravity-mediated
effects, although subdominant, are larger than Yukawa-generated
renormalization group (RG) effects and predict significant flavor
mixing for sleptons and sneutrinos, with potentially striking
implications for supersymmetric signals at the LHC.  In these models,
the lightest supersymmetric particle (LSP) is the gravitino, and the
next-to-lightest supersymmetric particle (NLSP) is metastable.  We
will examine a model in which the NLSP is a charged slepton, the
slepton mass splittings are large enough to be observable, and some
flavor mixings are significant.  Although we focus on this model, the
techniques developed are also applicable to models with other forms of
hybrid supersymmetry breaking (see, \eg,
Refs.~\cite{Chacko:2001km,Mohapatra:2008wx}) and other similar
frameworks.

With a metastable slepton NLSP, there is no missing energy associated
with the NLSP, and supersymmetric decay chains are, in principle,
fully reconstructible.  We choose a model in which the lightest two
sleptons are predominantly a selectron and a smuon.  These features
are favorable for our analysis.  Other features of the model are not
as favorable: the two lightest sleptons are quasi-degenerate, with a
mass splitting of roughly 5 GeV; the mixing of these two sleptons is
small; and one of the remaining sleptons is an almost pure stau, so
that its decays always involve taus which are hard to reconstruct.

In this analysis, we will only use information from electrons and
muons, and assume that taus are not reconstructed at all.  Even with
this restrictive assumption, we find that one can extract quite a lot
of information, including precision measurements of five slepton
masses and one mixing angle, as well as ${\cal O}(1)$ estimates of the
remaining mixing angles.  We expect that these results may be
improved, perhaps significantly, in a full analysis optimized to
extract information from tau events also.

\section{The Model}
\label{sec:model}

We now specify the model precisely and explain its key features.  In
our model, sfermion masses receive dominant flavor-conserving
contributions from gauge-mediated supersymmetry breaking (GMSB) and
smaller, but still significant, flavor-violating contributions from
gravity-mediated supersymmetry breaking.  Note that the
gravity-mediated contributions are generically present, since they
arise from generic K\"ahler potential contact terms between the
supersymmetry-breaking sector and the minimal supersymmetric standard
model (MSSM).

The size of the gravity-mediated contributions relative to the GMSB
contributions is
\begin{equation}\label{x1}
x \sim \frac{\tilde{m}_{\text{grav}}^2}{\tilde{m}_{\text{GMSB}}^2}
\sim \frac{1}{\nmess}
\left( \frac{\pi \mmess F_{\text{grav}}}{\alpha M_{\text{Pl}}
F_{\text{GMSB}}}\right)^2 \ ,
\end{equation}
where $\nmess$ is the number of $5 +\bar{5}$ messengers, $\mmess$ is
the messenger scale, $M_{\text{Pl}}$ is the Planck scale, and $\alpha$
stands for either the SU(2) or U(1) fine-structure constant.
$F_{\text{GMSB}}$ is the supersymmetry-breaking $F$-term associated
with gauge-mediation.  In addition, the supersymmetry-breaking sector
typically contains other $F$ terms as well, and the gravity-mediated
contributions are determined by $F_{\text{grav}}^2 = F_{\text{GMSB}}^2
+ \sum F_i^2$, where $F_i$ are the non-GMSB $F$-terms.  If these other
$F_i$ are negligible compared to $F_{\text{GMSB}}$, \eqref{x1} reduces
to
\begin{equation}
x \sim \frac{\tilde{m}_{\text{grav}}^2}{\tilde{m}_{\text{GMSB}}^2}
\sim \frac{1}{\nmess} 
\left( \frac{\pi \mmess}{\alpha M_{\text{Pl}}} \right)^2 \ .
\end{equation}
For the gravity contributions to be significant, either $\mmess$ must
be not far below $\alpha M_{\text{Pl}}$, or some $F_i$'s must be
larger than $F_{\text{GMSB}}$.
  
In the models we consider, we choose $x<1$, so that the lightest
supersymmetric particle (LSP) is the gravitino, with mass
$m_{\tilde{G}} \sim \sqrt{x} \, m_{\text{NLSP}}$.  The NLSP will be
generically long-lived in these models even for a low messenger scale,
since then its decays to the gravitino are suppressed by the large
$F_{\text{grav}}$.

The identity of the NLSP in these models is determined by the number
of messenger fields $\nmess$. For small $\nmess$ it is a neutralino,
and for large $\nmess$ it is a charged slepton.  We would like the
NLSP to be a charged slepton, with a large enough mass splitting
between the lightest neutralino and the NLSP that neutralino decays to
the NLSP are observable.  For very high messenger scales, however,
this requires a very large $\nmess$~\cite{Feng:1997zr}.  As an
alternative, we therefore choose a moderate messenger scale $\mmess
\sim 10^6~\gev$, and assume additional $F_i$'s so that $F_{\text{GMSB}}
\ll F_{\text{grav}}$.

For the gravity-mediated contributions, the constraints from
low-energy flavor-changing neutral current decays, such as $\mu\to
e\gamma$, imply that large slepton mixing is possible only if there is
strong degeneracy, while large mass splittings require a rather
precise lepton-slepton alignment.  We assume that the Froggatt-Nielsen
(FN) mechanism~\cite{Froggatt:1978nt} governs the flavor structure of
the gravity-mediated contributions~\cite{Nir:1993mx}.  As has been
shown in Ref.~\cite{Feng:2007ke}, there are many possible scenarios
that satisfy these constraints.  We are particularly interested in
models where the mass splitting between slepton generations is at
least a few GeV, so that ATLAS/CMS will perhaps be able to measure it,
and where at least some slepton mass eigenstates have appreciable
components of both electron and muon flavors, so that ATLAS/CMS will
have the potential of observing mixing.

Model B of Ref.~\cite{Feng:2007ke} is optimal for our purposes.
First, it has $x \sim 0.1$, so that the slepton mass splittings, which
are generically $\sqrt{x}$ times the slepton mass, can easily be
larger than 10 GeV.  Second, the flavor mixing, determined by a
$\text{U(1)} \times \text{U(1)}$ horizontal symmetry, is significant.
Schematically, the $(e,\mu,\tau)$ flavor decompositions of the slepton
mass eigenstates are
\begin{eqnarray}
\slepton_{6} &\sim& (\lambda^4,1,1) \no \\
\slepton_{5} &\sim& (\lambda^4,1,1) \no \\
\slepton_{4} &\sim& (1,\lambda^4,\lambda^4) \no \\
\slepton_{3} &\sim& (\lambda^4,\lambda^2,1) \no \\
\slepton_{2} &\sim& (\lambda^2,1,\lambda^2) \no \\
\slepton_{1} &\sim& (1,\lambda^2,\lambda^4) \ ,
\label{flavorlambda}
\end{eqnarray}
where $\lambda\sim0.2$ is the small breaking parameter of the FN
symmetry, and the sleptons are ordered by their masses from
$\slepton_6$, the heaviest, to $\slepton_1$, the lightest.\footnote{We
assume that the spurion $\lambda$ does not have an appreciable
$F$-term.  This may require a rather complicated FN
model~\cite{Ross:2002mr}.}  We see that there is significant mixing in
many, although not all, of the sleptons.  To fully specify Model B,
one must also choose 21 ${\cal O}(1)$ coefficients, one for each
independent component of the gravity-mediated contributions.  We use
this freedom to fix the mass ordering so that the lighter sleptons are
dominantly $e$- and $\mu$-flavored, since this improves the prospects
for precision mass and mixing measurements.

Our model, then, is completely specified by the input parameters for
the GMSB ``spine,'' the horizontal symmetry and charge assignments of
Model B, and the ${\cal O}(1)$ parameters entering the
gravity-mediated sfermion mass contributions.  These input parameters
are listed in \appref{details}.  We then give these input parameters
to \textsc{Spice}~\cite{Engelhard:2009br}, a computer program that
determines from these parameters the masses and flavor composition of
all superparticles, along with their flavor-general decay branching
ratios.  \textsc{Spice} interfaces with
\textsc{Softsusy}~\cite{Allanach:2001kg,Allanach:2009bv} to generate
the sparticle mass spectrum, and it calls
\textsc{Susyhit}~\cite{Djouadi:2006bz} to generate non-flavor
violating decays.  The decays generated by \textsc{Spice} include all
possible lepton flavor combinations for the decays of charginos,
neutralinos, sleptons, and sneutrinos.  All possible two-body decays
are included, along with relevant three-body slepton decays $\slepton
\to \slepton \ell \ell$ in cases where the only kinematically allowed
two-body decays involve gravitinos~\cite{Feng:2009bs}.  Note that,
because of flavor mixing, the sleptons and leptons in these 3-body
decays can, in principle, carry any flavor.  \textsc{Spice} produces
\textsc{Herwig}~\cite{Marchesini:1991ch,Corcella:2000bw} and SUSY Les
Houches Accord~\cite{Skands:2003cj} input files, which may be used to
generate collider events.  In this work, we use \textsc{Spice} with
\textsc{Softsusy} 3.0.2 and \textsc{Susyhit} 1.3.

\begin{figure}[tbp]
\includegraphics[width=0.7\textwidth]{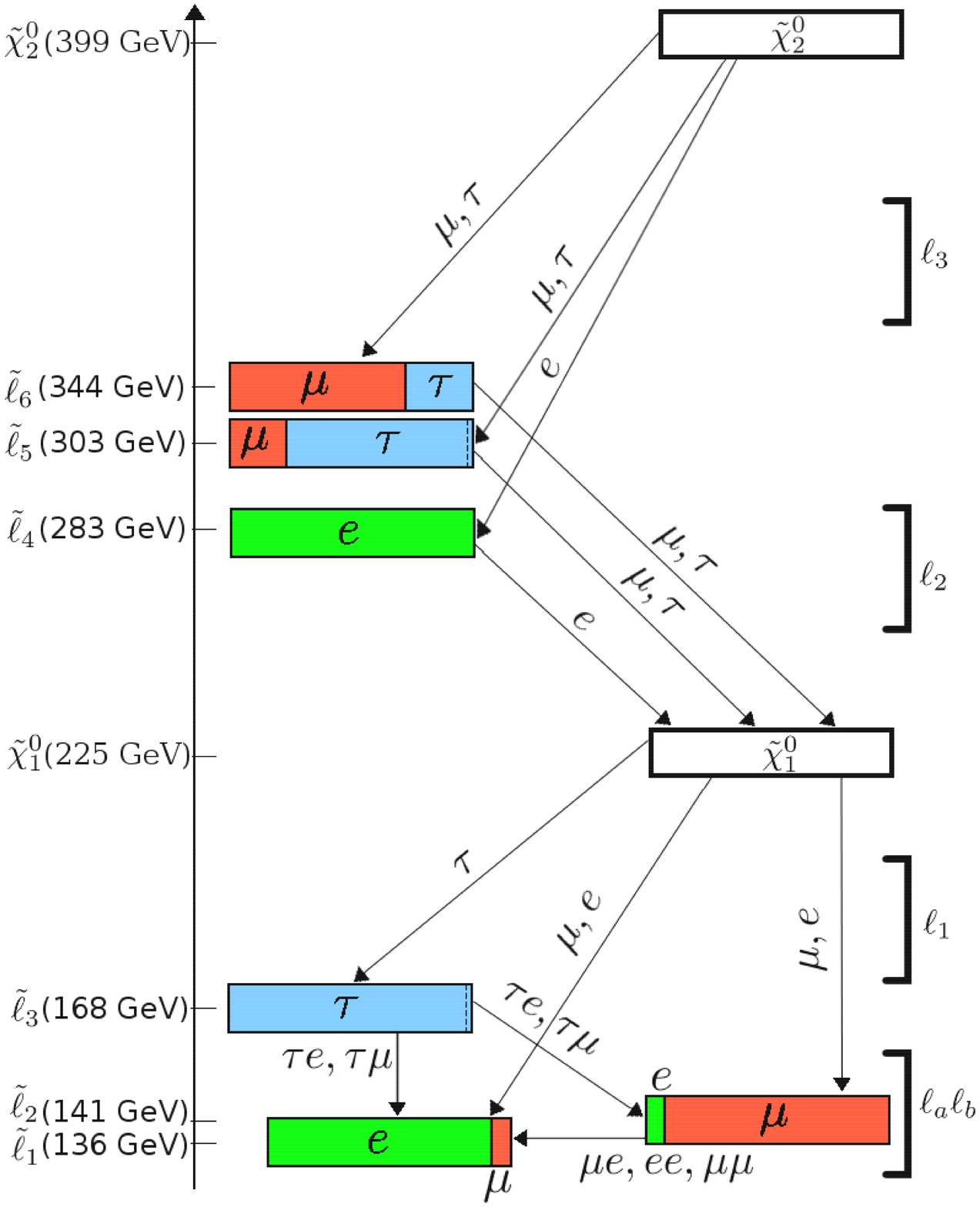}
\caption{\textit{Slepton Mass Hierarchy.}  Masses and flavor
compositions of the sleptons, along with the two lightest
neutralinos and the leptons resulting from common decay modes.
Sleptons $\slepton_{1,2,3}$ are dominantly right-handed, with the
dotted line in $\slepton_{3}$ demarcating a small $\tilde{\tau}_L$
component, and $\slepton_{4,5,6}$ are dominantly left-handed, with
the dotted line in $\slepton_{5}$ demarcating a small
$\tilde{\tau}_R$ component.  Left-right mixing in the other sleptons
is negligible.  The symbols $\ell_{1,2,3}$ and $\ell_{a,b}$ are used
to refer to leptons produced at particular steps in the decay chain,
regardless of their flavor.}
\label{fig:spectrum}
\end{figure}

The masses and flavor compositions of the six sleptons are given
graphically in \figref{spectrum}.  (Precise values for these masses
and mixings, and also for the sneutrinos, are also given in
\tablesref{sleptons}{sneutrinos}.)  Although \textsc{Spice} properly
accounts for full $6 \times 6$ slepton mixing, in this model there are
no $A$-term contributions associated with the gravity-mediated
contributions, and so the only left-right mixing is small and
originates from RG evolution.  Thus, the lighter 3 sleptons are
dominantly right-handed and the heavier 3 sleptons are dominantly
left-handed, with flavor compositions consistent with
\eqref{flavorlambda}.  Left-right mixing is small, and limited to the
stau-dominated states $\slepton_3$ and $\slepton_5$.  As desired,
$\slepton_1$ and $\slepton_2$ have some $\tilde{e}_R-\tilde{\mu}_R$
mixing and $\tilde{\tau}_R$ is confined primarily to
$\slepton_3$.  The state $\slepton_4$ is dominantly $\tilde{e}_L$,
while $\slepton_{5,6}$ demonstrate ${\cal O}(1)$
$\tilde{\mu}_L-\tilde{\tau}_L$ mixing.

As shown in \figref{spectrum}, the lowest lying supersymmetric states
have the following hierarchy:
\begin{equation}
m_{\tilde G}\ll m_{\slepton_{1,2,3}} < m_{\chi_1^0} <
m_{\slepton_{4,5,6}} < m_{\chi_2^0} \ .
\end{equation}
Here $\chi_1^0$ is the lightest neutralino, which is mostly the Bino,
and $\chi_2^0$ is the second-lightest neutralino, which is, to a good
approximation, the neutral Wino.  The dominant decay modes are
\begin{eqnarray}
\chi^0_2 &\to& {\ell}_3^{\pm} \slepton_{4,5,6}^{\mp} \\
\slepton_{4,5,6}^{\pm} &\to& {\ell}_2^{\pm} \chi^0_1 \\
\chi^0_1 &\to& {\ell}_1^{\pm} \slepton_{1,2,3}^{\mp} \\
\slepton_{3}^\pm &\to& {\ell}_a^\pm {\ell}_b^\mp \slepton_{1,2}^\pm ,
{\ell}_a^\pm {\ell}_b^\pm \slepton_{1,2}^\mp \\
\slepton_{2}^\pm &\to& {\ell}_a^\pm {\ell}_b^\mp \slepton_1^\pm ,
{\ell}_a^\pm {\ell}_b^\pm \slepton_1^\mp \ .
\end{eqnarray}
These are also shown in \figref{spectrum}.  There are several
noteworthy features.  First, the decays of $\chi^0_2$ directly to the
light sleptons $\slepton_{1,2,3}$ are highly suppressed, since these
sleptons are dominantly right-handed and have no couplings to Winos.
Second, the leptons in the $\slepton_{2,3}$ decays are quite soft and
may be difficult to detect.  Finally, the slepton-to-slepton
$\slepton_{2,3}$ decays may be either charge-preserving or
charge-flipping.  The charge-flipping modes are possible because the
neutralino is a Majorana fermion, and the possibility of both kinds of
decays will play a crucial role in the analysis described below.

\section{Event Generation}
\label{sec:event}

We generate our signal events for the LHC with center-of-mass energy
$\sqrt{s} = 14~\tev$ using
\textsc{Herwig}~\cite{Marchesini:1991ch,Corcella:2000bw} and pass
them through a generic LHC detector simulation, \textsc{AcerDET
1.0}~\cite{RichterWas:2002ch}. We configure \textsc{AcerDET} as
follows: electrons and muons are selected with $p_T>6$ GeV and
$|\eta|<2.5$. Electrons and muons are considered to be isolated if
they are separated from other leptons and jets by $\Delta R>0.4$,
where $\Delta R = \sqrt{(\Delta\eta)^2+(\Delta\phi)^2}\, $, and if
less than 10 GeV of energy is deposited in a cone of $\Delta R =
0.2$. The lepton momentum resolutions we use are parameterized from
the results of Full Simulation of the ATLAS
detector~\cite{Aad:2008zzm}; our electrons are smeared according to a
pseudorapidity-dependent parametrization, while muons are smeared
according to the results for $|\eta|<1.1$.  The leading-order
two-to-two supersymmetric cross section determined by \textsc{Herwig}
with CTEQ5L parton distribution functions is 1.154 pb.  We generate
115,400~events, corresponding to $100~\ifb$ of data.  \textsc{AcerDET}
does not take into account lepton reconstruction efficiencies. We
therefore apply by hand a reconstruction efficiency of 90\% to the
muons and a reconstruction efficiency of 77\% to the electrons. This
gives 0.86 as the ratio of electron to muon reconstruction efficiency.

In a real detector or full simulation thereof, a long-lived charged
slepton, such as our $\slepton_{1}$, would be expected to produce a
visible track by virtue of its charge.  Measurement of the curvature
of this track would determine the $\slepton_{1}$ {\em
momentum}.\footnote{Strictly speaking, the curvature determines only
the momentum-to-charge ratio, but we will assume unit charges for all
heavy charged tracks.}  If the $\slepton_{1}$ arrival time were also
to be measured with sufficient accuracy, the $\slepton_{1}$ {\em
speed} could also be determined. The momentum and speed may be
combined to determine the track {\em mass}, which will constitute the
primary signature for the existence of new physics (\ie, the
$\slepton_{1}$) in our signal events.  \textsc{AcerDET} is not a full
detector simulation and so, unsurprisingly, does not produce track
objects or speed ``measurements'' associated with our long-lived
sleptons.

Since measurement errors associated with the $\slepton_{1}$ are
potentially very important in our study, it is necessary to augment
\textsc{AcerDET} by incorporating additional parameterizations for
$\slepton_{1}$ momentum and speed measurements.  We used the
resolutions taken from Ref.~\cite{Ellis:2006vu}, based on ATLAS Muon
Resistive Plate Chamber (RPC) timings.  Given these results, to model
the reconstructed $\slepton_{1}$ momentum, we start with the momentum
taken from Monte Carlo truth, and then smear the slepton's 3-momentum
magnitudes $p \equiv | \vec{p}_{\slepton_1}|$ and speeds $\beta$ by
Gaussian distributions with $\sigma_p/p = 0.05$ and $\sigma_\beta =
0.02$, respectively.  The slepton's energy and direction are not
smeared. Note that we could just as easily have used another
parametrization based on ATLAS Muon Drift Tube (MDT)
fits~\cite{Tarem:2009zz}.  Our use of just one such source of timing
information might therefore mean that our speed resolution can be
improved in a future study.  We only consider sleptons with $\beta >
0.6$ to be reconstructible, because sleptons moving any slower are
unlikely to reach the muon chambers quickly enough to be registered in
the same bunch crossing. In our events, 95$\%$ of sleptons have a
speed, $\beta$, greater than 0.6.

Having parametrized the slepton reconstruction process above,
subsequent uses of expressions like ``$\slepton_{1}$ momentum'' and
``$\beta$'' refer to the smeared (supposedly reconstructed) quantities
rather than to Monte Carlo truth.

\section{Analysis}
\label{sec:analysis}

In this section we outline our approach to measuring the mixings and
masses of the six sleptons. As our NLSP is a metastable charged
particle, our entire decay chain is, in principle, fully
reconstructible. Our goal is therefore to reconstruct the various
superpartners, starting with the slepton NLSP, by working our way up
the decay chain, constructing various invariant mass distributions. In
this section we explain each invariant mass distribution and in the
next section we interpret the results in terms of mixings and masses.

As explained in \secref{intro}, in this analysis we ignore tau
leptons, and instead only consider what we can deduce about masses and
mixings from observing electrons and muons.  We make this choice
because tau reconstruction is expected to be poor in comparison to
electron and muon reconstruction, but the inclusion of tau
reconstruction certainly merits further study.

\subsection{Reconstructing the slepton NLSP {\boldmath$\slepton_1$}}

In each event, direct reconstruction of the $\slepton_{1}$ momentum,
speed, and mass is expected to be relatively straightforward, using
the slow charged track signature already described in \secref{event}.
We restrict our attention to sleptons with speeds in the range $0.6 <
\beta < 0.8$.  The necessity of the lower bound has already been
mentioned.  The upper bound effectively distinguishes supersymmetric
events from
background~\cite{Feng:2005ba,Rajaraman:2006mr,Rajaraman:2007ae,CMSNote},
and also plays a crucial role in improving the resolution of the
$\slepton_{1}$ mass measurement. Of the sleptons with $\beta>0.6$,
$15\%$ have $\beta<0.8$ and so pass this extra restriction. The
reconstructed mass distribution is shown in \figref{slepton1} before
and after we impose the upper $\beta$ limit. As expected, even
ignoring background that will enter the $\beta > 0.6$ distribution
(but is not included in \figref{slepton1}a), this distribution is fit
poorly by a Gaussian curve, and hence we obtain a better measurement
of slepton mass by considering the slowest of the sleptons. After
requiring $0.6 < \beta < 0.8$, we assume that the efficiency for
reconstructing $\slepton_1$ is 100\%, and we do not require these
sleptons to pass any isolation criteria.

\begin{figure}[tbp]
\begin{center}
\subfigure[$\beta>0.6$]{
\includegraphics[width=0.44\textwidth,clip]{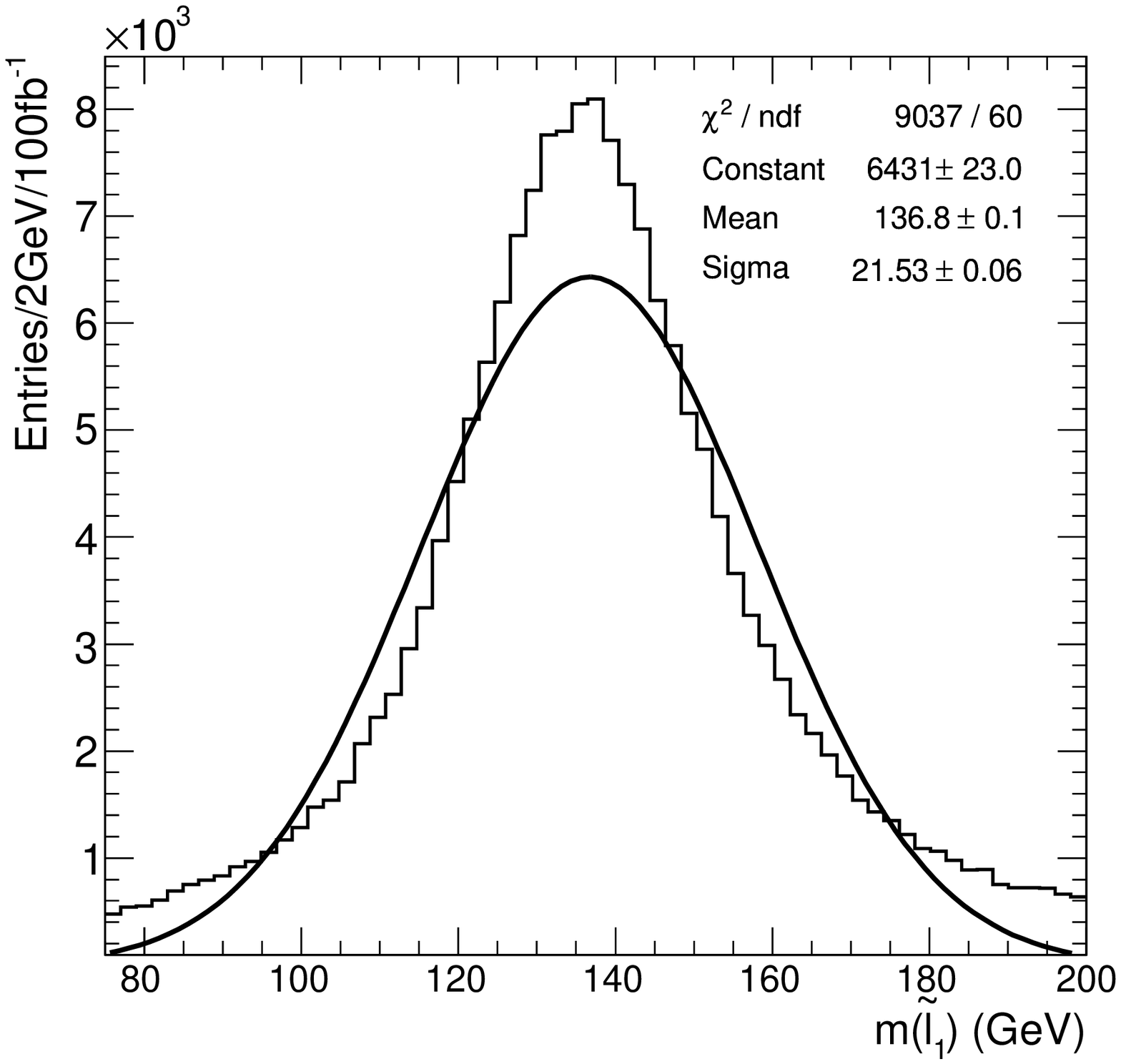}\vspace*{-.3in}
}
\subfigure[$0.6<\beta<0.8$]{
\includegraphics[width=0.44\textwidth,clip]{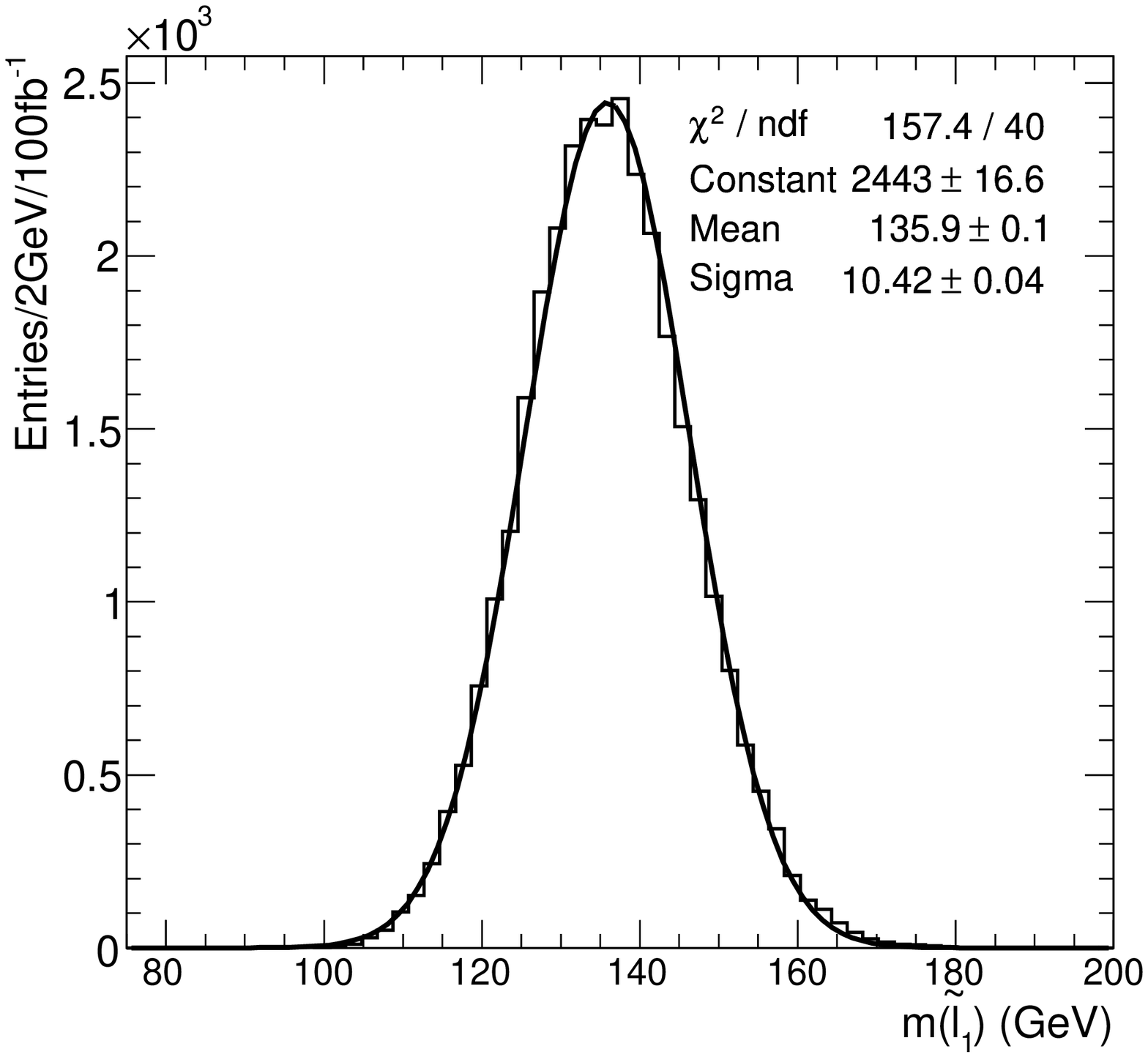}\vspace*{-.3in}
}
\vspace{-.1in}
\caption{\label{fig:slepton1} \textit{Reconstructed $\slepton_1$ Mass
    Distributions.}  Reconstructed $\slepton_{1}$ masses from events
    with slepton speed (a) $\beta>0.6$ and (b) $0.6<\beta<0.8$.  The
    histograms are the distributions, and the solid lines are Gaussian
    fits with means and standard deviations as indicated.  The
    Gaussian fit for (a) is poor and is shown for comparison purposes
    only, as described in the text.  The $\slepton_{1}$'s 3-momentum
    magnitudes $p \equiv | \vec{p}_{\slepton_1}|$ and speeds have been
    smeared by Gaussian distributions with $\sigma_p/p = 0.05$ and
    $\sigma_\beta = 0.02$, respectively.}
\end{center}
\end{figure}

\Figref{slepton1} tells us two valuable things.  First, for any given
event, the slepton mass is measured to an accuracy of the order of 10
GeV.  Second, from taking all of the events together, the {\em
statistical} error on the mean mass is much smaller, of the order of
0.1 GeV.  The strong agreement between the underlying $\slepton_1$
mass and the fitted mean reconstructed $\slepton_1$ mass in
\figref{slepton1} is perhaps misleading.  Such agreement was to be
expected as no sources of systematic offsets were introduced in our
smearing process. Realistically, the fitted mean reconstructed
$\slepton_1$ mass will be subject to some systematic offset, even
after all attempts to calibrate the detector have been completed. The
size and direction of such an offset cannot be known in advance, so we
do not attempt to simulate any systematic offset in this analysis.

We can make use of the high precision (mean) mass measurement from the
set of all events to help us remove some of the momentum and
time-of-flight measurement errors in individual events, thereby
reducing our exposure to the 10 GeV event-by-event variability in
reconstructed slepton masses.  We do this by scaling each component of
the four-momentum of every slow charged track by a constant so that
the track mass matches the mean mass obtained from the fit to all
events.  After rescaling, the smeared and rescaled momentum is
centered on the true momentum with a full-width half-maximum of $27$
GeV and rms of 9.9 GeV, and the corresponding energy difference
distribution has a full-width half-maximum of $31$ GeV and rms of 9.4
GeV.  We find that this event-by-event rescaling process is necessary
to allow us sufficient resolution to travel up the decay chain and
determine the masses of sparticles heavier than the NLSP.

\subsection{Why it is impossible to directly reconstruct 
{\boldmath $\slepton_{2,3}$}}

The next most obvious particles to reconstruct are the next two
lightest sleptons, $\slepton_{2,3}$. We see in \figref{spectrum} that
the dominant $\slepton_{2,3}$ decays are three-body. In principle,
given an ideal detector, we could reconstruct the $\slepton_{2,3}$ by
looking at the three-particle invariant mass distributions resulting
from combining the $\slepton_{1}$ with all possible combinations of
two further leptons which give a charged $\slepton_{2,3}$ candidate.

Since in this analysis we are not reconstructing taus, we cannot
detect $\slepton_3$ in this way.  We will briefly comment on the
$\slepton_3$ in \secref{slepton3}.

Unfortunately, direct reconstruction of $\slepton_{2}$ is also
impossible as $\slepton_{1}$ and $\slepton_{2}$ are nearly
degenerate. In \figref{leptons} we plot the true $p_T$ of all leptons
produced in the three-body $\slepton_{2}$ decays. We see that $\sim
90\%$ of all electrons and muons have $p_T < 10~\gev$. Leptons with
$p_T < 10~\gev$, which we denote ``soft'' leptons, will be very
difficult to reconstruct in ATLAS/CMS.

\begin{figure}[tbp]
\begin{center}
\includegraphics[width=0.5\textwidth]{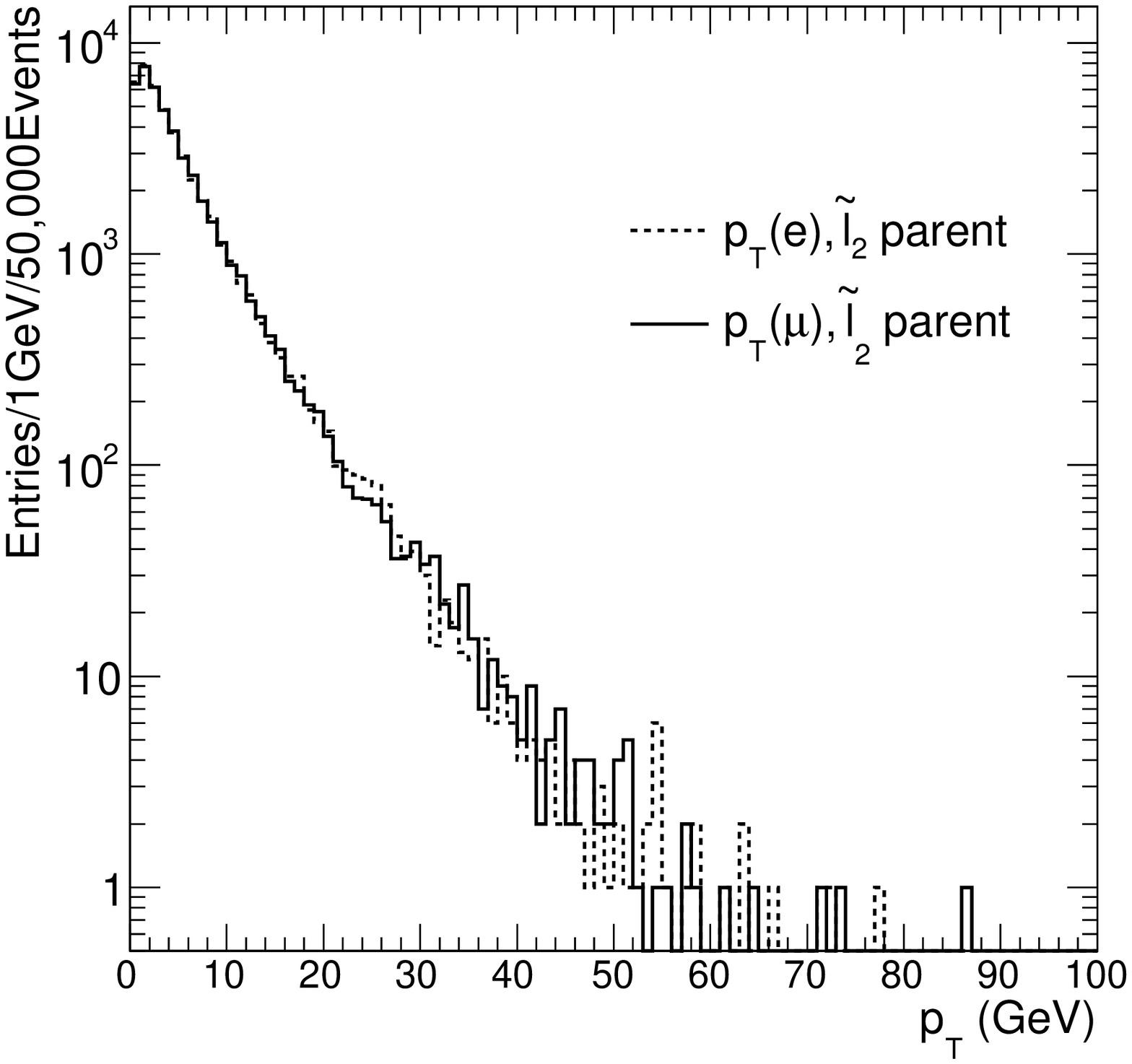}
\caption{\label{fig:leptons} \textit{Three Body $p_T$ Distribution.}
The $p_T$ distribution of leptons $\ell = e, \mu$ from the three
body decays $\slepton_2 \to \ell \ell \slepton_1$ from Monte Carlo
truth.}
\end{center}
\end{figure}

To determine the mass and mixings of the $\slepton_{2}$, we must
therefore rely on indirect measurements. These do, in fact, make a
measurement of the $\slepton_2$ mass possible, as we describe in the
next section.

\subsection{Reconstructing the lightest neutralino {\boldmath $\chi^0_1$} and
{\boldmath$\slepton_2$}}

In this section we describe the $\chi^0_1$ reconstruction, and explain
how it can be used to detect $\slepton_2$ and measure its mass as
well.

Referring to \figref{spectrum}, we see that the neutralino can decay
to any one of the three light sleptons.  Let us first discuss
neutralino decays to $\slepton_1$ and $\slepton_2$, with the
$\slepton_2$ subsequently decaying to $\slepton_1$ via a 3-body decay.
Since the leptons involved in this 3-body decay, $\ell_{a,b}$, are
typically soft, they will usually go undetected, so that one only
observes the final $\slepton_1$ and the hard lepton, $\ell_1$, coming
from the original neutralino decay.  (Throughout this section,
subscripts on the leptons are as shown in \figref{spectrum}.)  We
therefore define two main categories of neutralino decay --- direct
and indirect. The latter is further divided into two subcategories,
which we name ``OS indirect'' and ``SS indirect,'' based on whether
the final $\slepton_1$ has the opposite or same sign as the lepton
$\ell_1$:
\begin{itemize}
\item Direct decays of $\chi^{0}_{1}$ to the slepton NLSP: 
$\chi^0_1\to\slepton^{\pm}_{1}{\ell}^{\mp}_1$
\item Indirect decays of the $\chi^{0}_{1}$ to the slepton NLSP via
$\slepton_2$,
    $\chi^0_1\to\slepton^{\pm}_{2}{\ell}^{\mp}_1$, followed by
    a three-body $\slepton_{2}$ decay of one of the following types:
    \begin{itemize}
    \item[$\circ$]
      $\slepton^{\pm}_{2}\to\slepton^{\pm}_{1}{\ell}^\pm_a
{\ell}^\mp_b$ \quad OS indirect
    \item[$\circ$]
      $\slepton^{\pm}_{2}\to\slepton^{\mp}_{1}{\ell}^\pm_a
{\ell}^\pm_b$ \quad SS indirect
  \end{itemize}
\end{itemize}
Note that in the instance of a direct decay, the $\slepton_1$ and the
lepton $\ell_1$ must be oppositely charged, whereas in indirect decays,
same-sign charges are also possible.  Detection of same-sign events
are direct evidence of the Majorana nature of the neutralino.  As
mentioned above, with a hard lepton $p_T$ cut, only the lepton $\ell_1$
is observed.

\subsubsection{Direct and OS-indirect $\chi^0_1$ decays:
measuring the $\chi^0_1$ and $\slepton_2$ masses}

For both the direct and OS indirect decays, the $\slepton_{1}$ and the
lepton $\ell_1$ have opposite signs. These channels therefore have
identical signatures and are therefore reconstructed in tandem. We
take all OS $\slepton_1^\pm\, e^\mp$ and $\slepton_1^\pm\, \mu^\mp$
combinations and reconstruct the invariant masses $m_{\slepton_1^\pm
e^\mp}$ and $m_{\slepton_1^\pm\mu^\mp}$.

When our OS $\slepton_1$ and lepton are from a direct decay, we expect
to reconstruct the $\chi^0_1$ mass exactly.  However, when the
$\slepton_1$ results from an indirect decay, we do not expect to
correctly reconstruct the $\chi^0_1$ as we are missing the two soft
leptons produced in the decay $\slepton_2^\pm\to\slepton_1^\pm\,
{\ell}_a^+ {\ell}_b^-$.  Such events nevertheless contain valuable
information: as shown in Ref.~\cite{Feng:2009yq}, the
$\slepton_1$-lepton invariant mass then exhibits a ``shifted peak,''
somewhat lower than the neutralino mass, by an amount
\begin{equation} 
E_{\text{shift}} \simeq \frac{M^2+m_1^2}{2M m_1} \Delta m \ .
\label{eshift}
\end{equation}
Here $M$ is the neutralino mass, the mean of the $\slepton_1 \ell_1$
invariant mass distribution, $m_1$ is the reconstructed mean of the
$\slepton_1$ mass, and $\Delta m \equiv m_{\slepton_2} -
m_{\slepton_1}$.  For our model parameters we predict
$E_{\text{shift}} \simeq 5.6~\gev$. By measuring this shift we can
deduce $\Delta m$ and determine the mass of the $\slepton_{2}$
indirectly.

We plot these OS slepton-lepton invariant mass distributions in
\figref{OS_neutralino} for both $p_T > 10~\gev$ and $p_T >
30~\gev$. The harder $p_T$ cut effectively removes the soft leptons
produced in three-body $\slepton_{2,3}$ decays that have not already
failed reconstruction from the lepton collections (see
\figref{leptons}). We decompose these distributions (and the
distributions that follow) into the sum of an exponentially falling
background and one or more Gaussian peaks.  The exponentially falling
background is designed to model the combinatoric background from
supersymmetric events.  The fitting function has the form
\begin{equation}
\label{fit}
\frac{dN}{dm} = N_{\text{tot}}
\left[ \left( 1- \sum_i f_i \right) (-a_i) e^{a_i m}
+ \sum_i f_i \sqrt{\frac{2}{\pi}} \frac{1}{\sigma_i} 
e^{-\frac{(m - \bar{m}_i)^2 }{2\sigma_i^2}} \right] \ ,
\end{equation}
where $N_{\text{tot}}$ is the total number of events in the
distribution, $a_i$ is the exponential decay parameter with units of
$\gev^{-1}$, $f_i$ is the fraction of the total number of events in
peak $i$, and $\bar{m}_i$ and $\sigma_i$ are the center and width of
peak $i$.  

\begin{figure}[tbp]
\begin{center}
\subfigure[ $\ \ell = e$]{
\includegraphics[width=0.44\textwidth,clip]{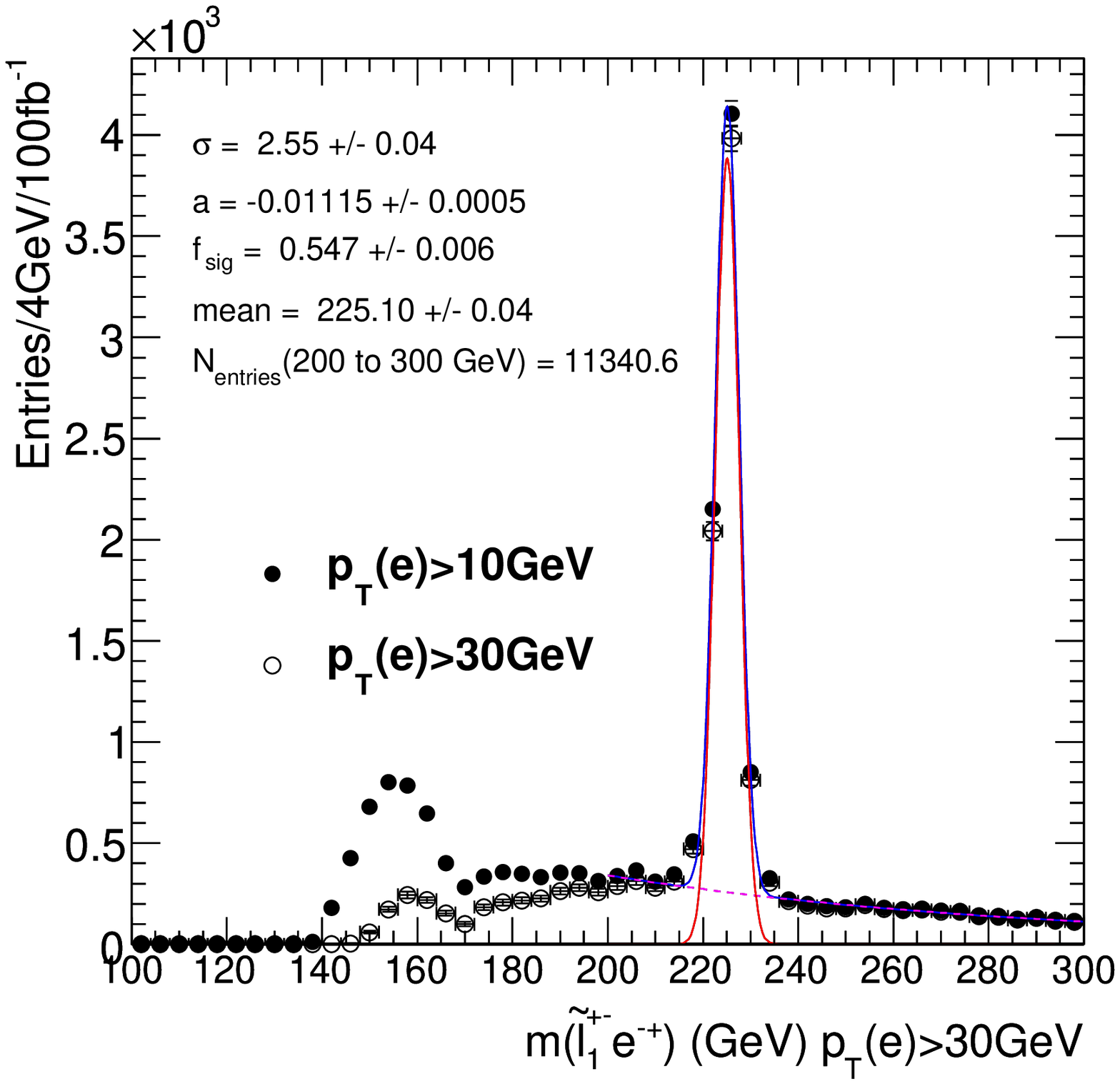}
}
\subfigure[ $\ \ell = \mu$]{
\includegraphics[width=0.44\textwidth,clip]{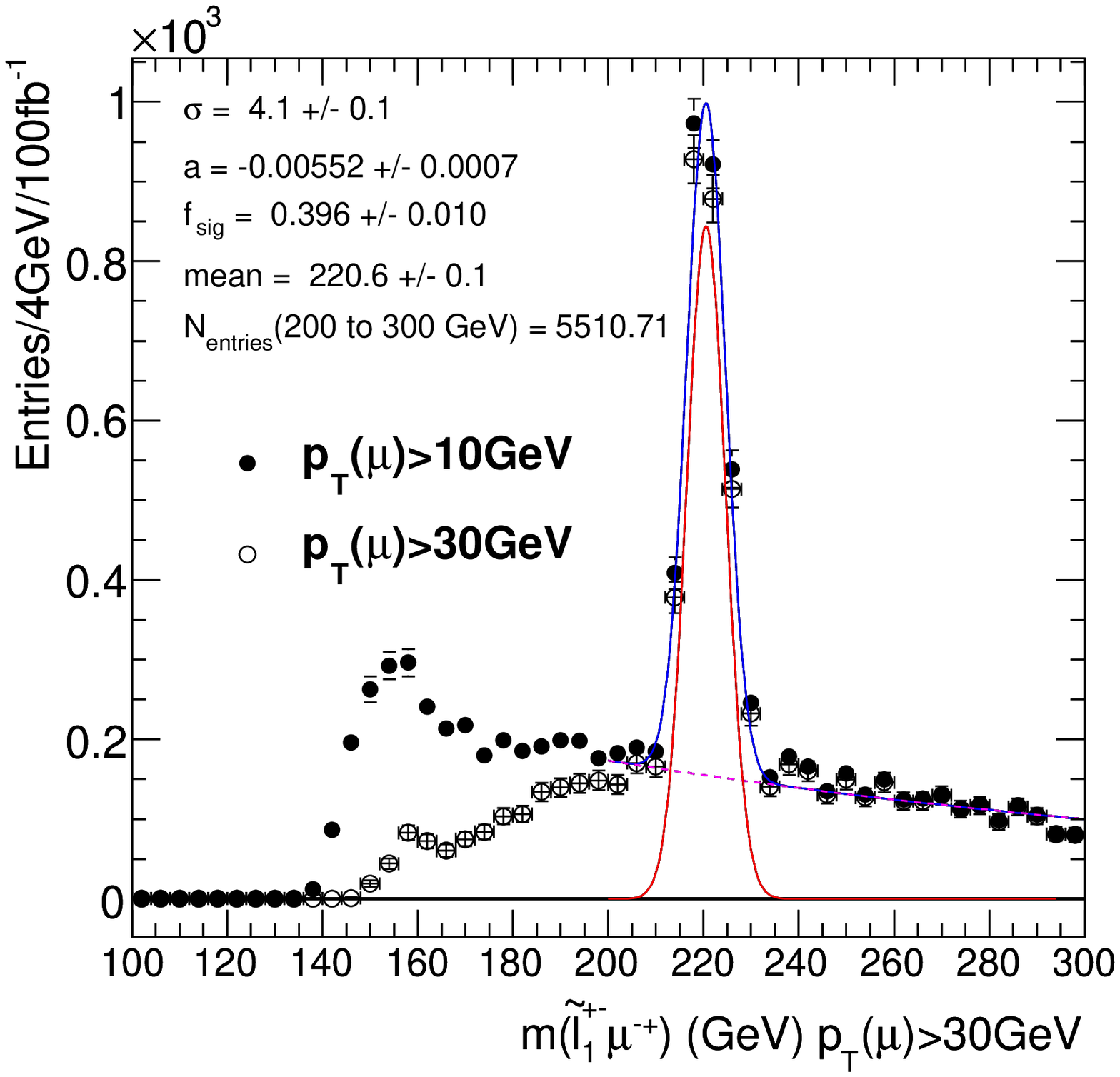}
}
\caption{\label{fig:OS_neutralino} \textit{OS $\slepton^{\pm}_{1}\,
    {\ell}^{\mp}$ Invariant Mass Distributions.}  Invariant mass
    distributions of OS $\slepton^{\pm}_{1}\, {\ell}^{\mp}$ pairs for
    (a)~$\ell = e$ and (b)~$\ell = \mu$, requiring $p_T > 10~\gev$ and
    $p_T > 30~\gev$.  The $p_T > 30~\gev$ distributions in the range
    $200~\gev < m < 300~\gev$ have been fit with a Gaussian peak on
    top of an exponentially decaying background as given by the red
    (lighter) solid and purple (lighter) dashed contours, respectively.
    The sum of these fits is given by the blue (darker) solid
    line. The fit parameters, defined in \eqref{fit}, are as
    indicated.  }
\end{center}
\end{figure}

Indeed, we see two different peaks: a higher one at 225.1~GeV in the
$\slepton_1^\pm e^\mp$ sample, corresponding to the true neutralino
mass, and the lower shifted peak centered at 220.6~GeV in the
$\slepton_1^\pm \mu^\mp$ sample.  Of course, in the model we are
considering, each one of these samples contains both these peaks, but
because of the flavor compositions of $\slepton_1$ and $\slepton_2$,
the $\slepton_1^\pm e^\mp$ sample is dominated by direct decays, and
therefore seems to exhibit just the unshifted peak at $\sim225$~GeV,
while the $\slepton_1^\pm \mu^\mp$ sample is dominated by indirect
decays, and therefore seems to exhibit just the shifted peak at
$\sim220$~GeV.  In a general model, with larger mixings, one would see
a double-peak structure, which might be harder to disentangle.  Then,
however, the SS decays can come to our aid: these can only originate
from indirect decays, and thus will cleanly exhibit only the shifted
peak, with no contamination from direct decays.

\subsubsection{SS indirect $\chi^0_1$ decays: a clean measurement
of the $\slepton_2$ mass}
\label{sec:sssection}

As explained above, the SS $\slepton_1 ^\pm \ell^\pm$ sample cleanly
probes the indirect neutralino decays through $\slepton_2$, and
exhibits just the neutralino shifted peak.  Here we take all SS
$\slepton_1^\pm\,e^\pm$ and $\slepton_1^\pm\,\mu^\pm$ combinations and
reconstruct the invariant masses $m_{\slepton_1^\pm e^\pm}$ and
$m_{\slepton_1^\pm\mu^\pm}$. Again we do so for both $p_T > 10~\gev$
and $p_T > 30~\gev$.  The invariant mass distributions are shown in
\figref{SS_neutralino}, and indeed, we only see the neutralino shifted
peak at around 219 GeV.  It is not surprising that the peaks in these
distributions are somewhat lower than the shifted peak of
\figref{OS_neutralino}b.  The SS samples necessarily come from
neutralino decays through $\slepton_2$, which only exhibit the shifted
peak, whereas the OS $\slepton_1\mu$ sample, although dominated by
such decays, also contains some events in which the neutralino decays
directly to $\slepton_1$.  The latter lead to the correct neutralino
peak at 225 GeV.

\begin{figure}[tbp]
\begin{center}
\subfigure[ $\ \ell = e$]{
\includegraphics[width=0.44\textwidth,clip]{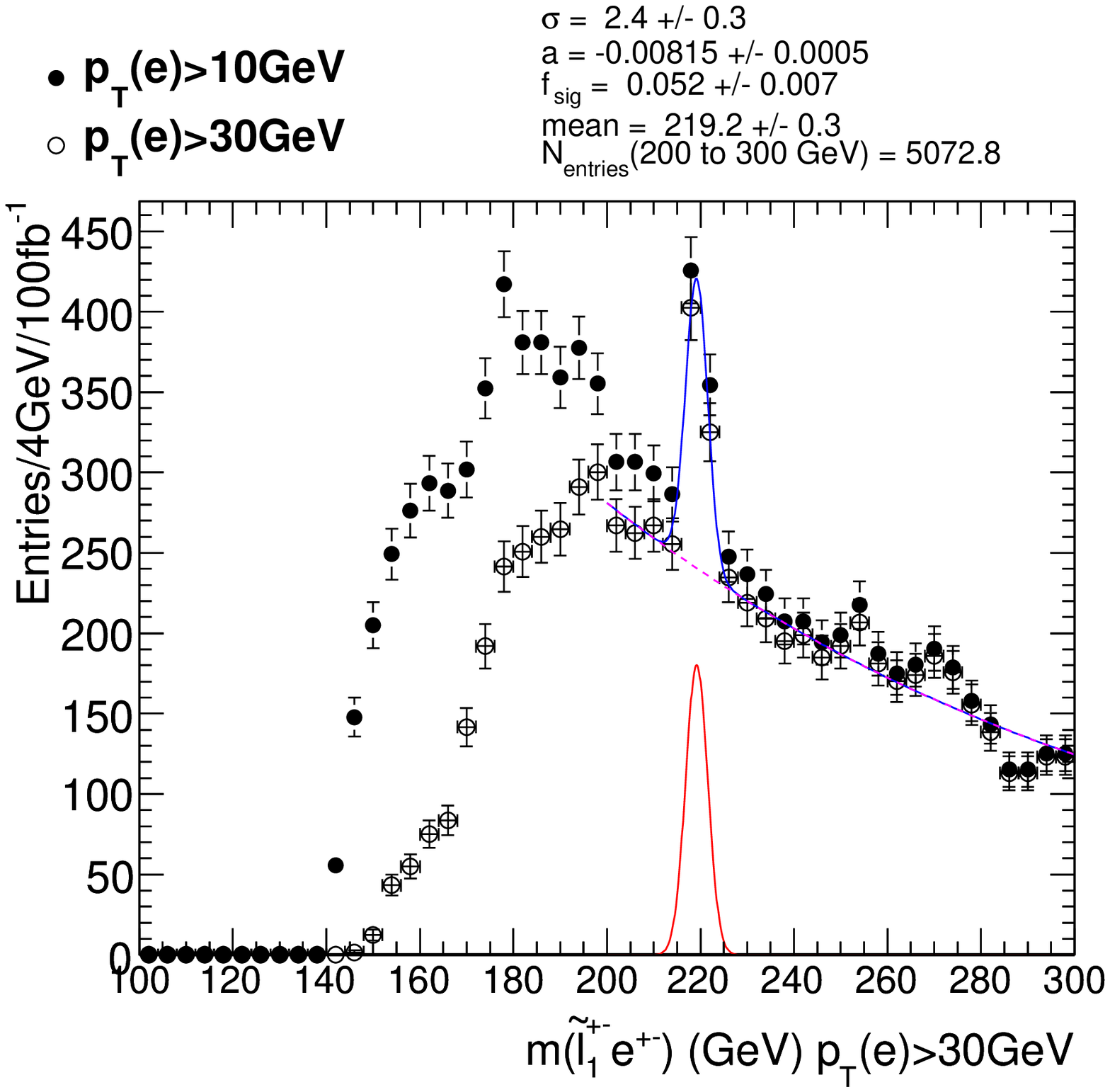}
}
\subfigure[ $\ \ell = \mu$]{
\includegraphics[width=0.44\textwidth,clip]{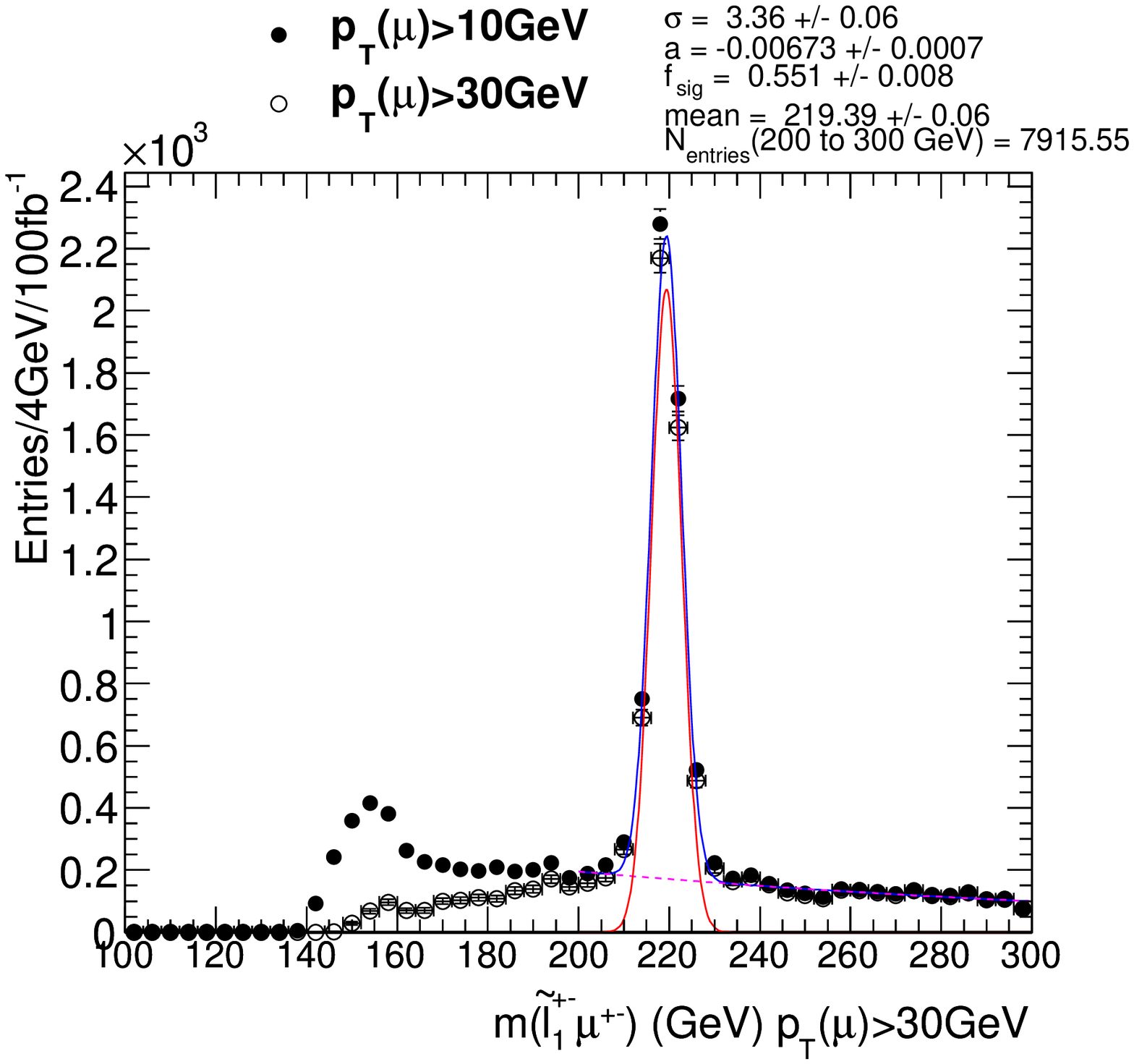}
}
\caption{\label{fig:SS_neutralino} \textit{SS $\slepton^{\pm}_{1}\,
{\ell}^{\pm}$ Invariant Mass Distributions.}  As in
\figref{OS_neutralino}, but for SS $\slepton^{\pm}_{1} {\ell}^{\pm}$
pairs.}
\end{center}
\end{figure}

\subsection{Indirectly reconstructing the remaining light slepton
{\boldmath$\slepton_3$}}
\label{sec:slepton3}

Inspecting \figsref{OS_neutralino}{SS_neutralino} also reveals a
smaller peak around 160 GeV.  It is easy to see that this peak
corresponds, at least partially, to $\slepton_3$.  Consider
$\slepton_3$ decays to $\slepton_1$.  These are dominantly
$\slepton_3\to\slepton_1 e \tau$.  The tau decay could give another
charged lepton, but this charged lepton would typically be softer than
the original electron produced in the $\slepton_3$ decay.  Thus, when
we consider the invariant mass of this electron paired with the
$\slepton_1$, we should find a peak somewhat below the $\slepton_3$
mass, much like the shifted neutralino peak discussed in the previous
two sections.  An analogous peak should also occur for $\slepton_1
\mu$ pairs, coming either from direct $\slepton_3 \to \slepton_1 \mu
\tau$ decays, or from indirect decays through $\slepton_3 \to
\slepton_2 \mu \tau$.

\begin{figure}[tbp]
\begin{center}
\subfigure[ $\ (\ell_1, \ell')=(e, \mu)$]{
\includegraphics[width=0.44\textwidth,clip]{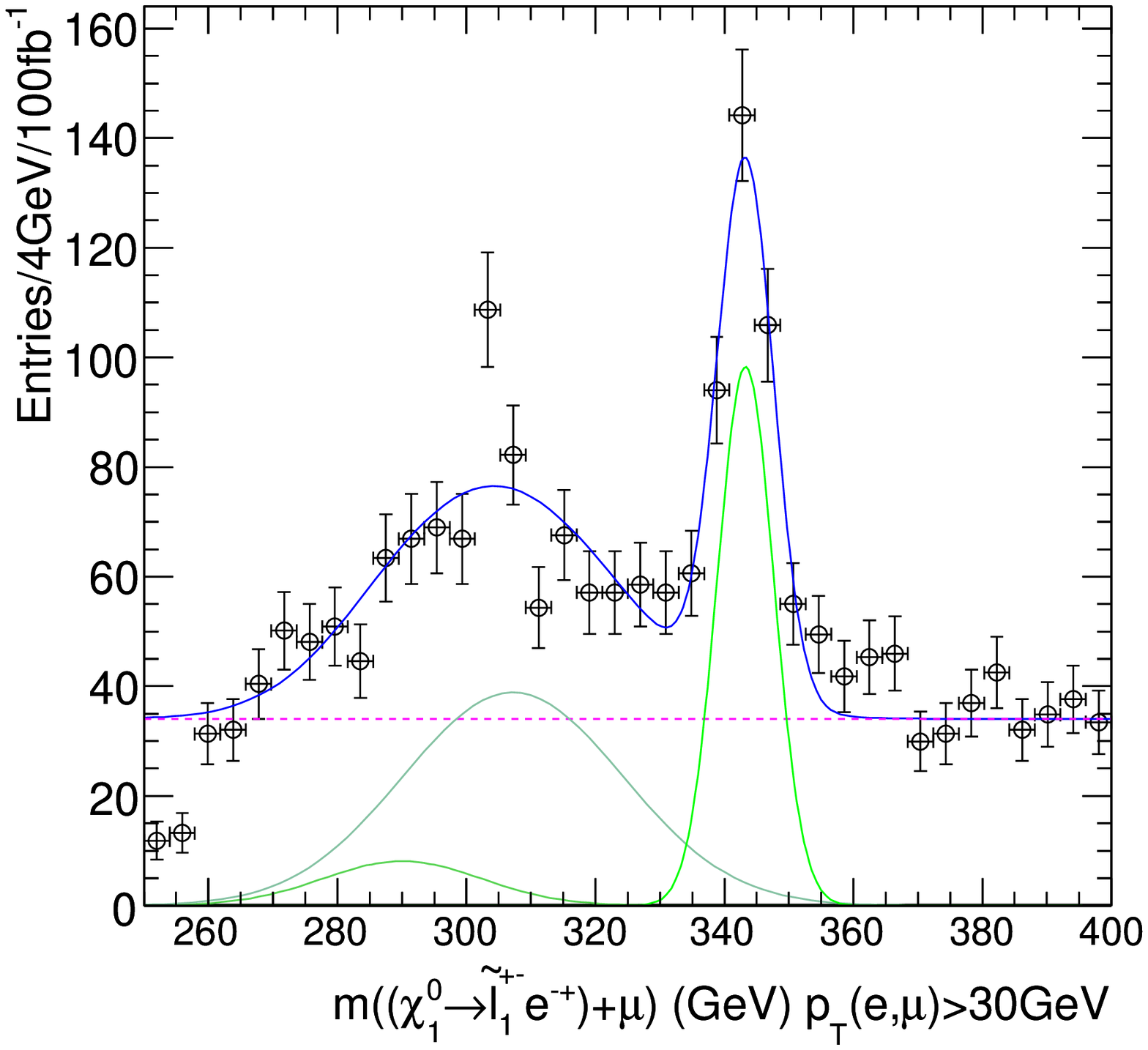} 
}
\subfigure[ $\ (\ell_1, \ell')=(e, e)$]{
\includegraphics[width=0.44\textwidth,clip]{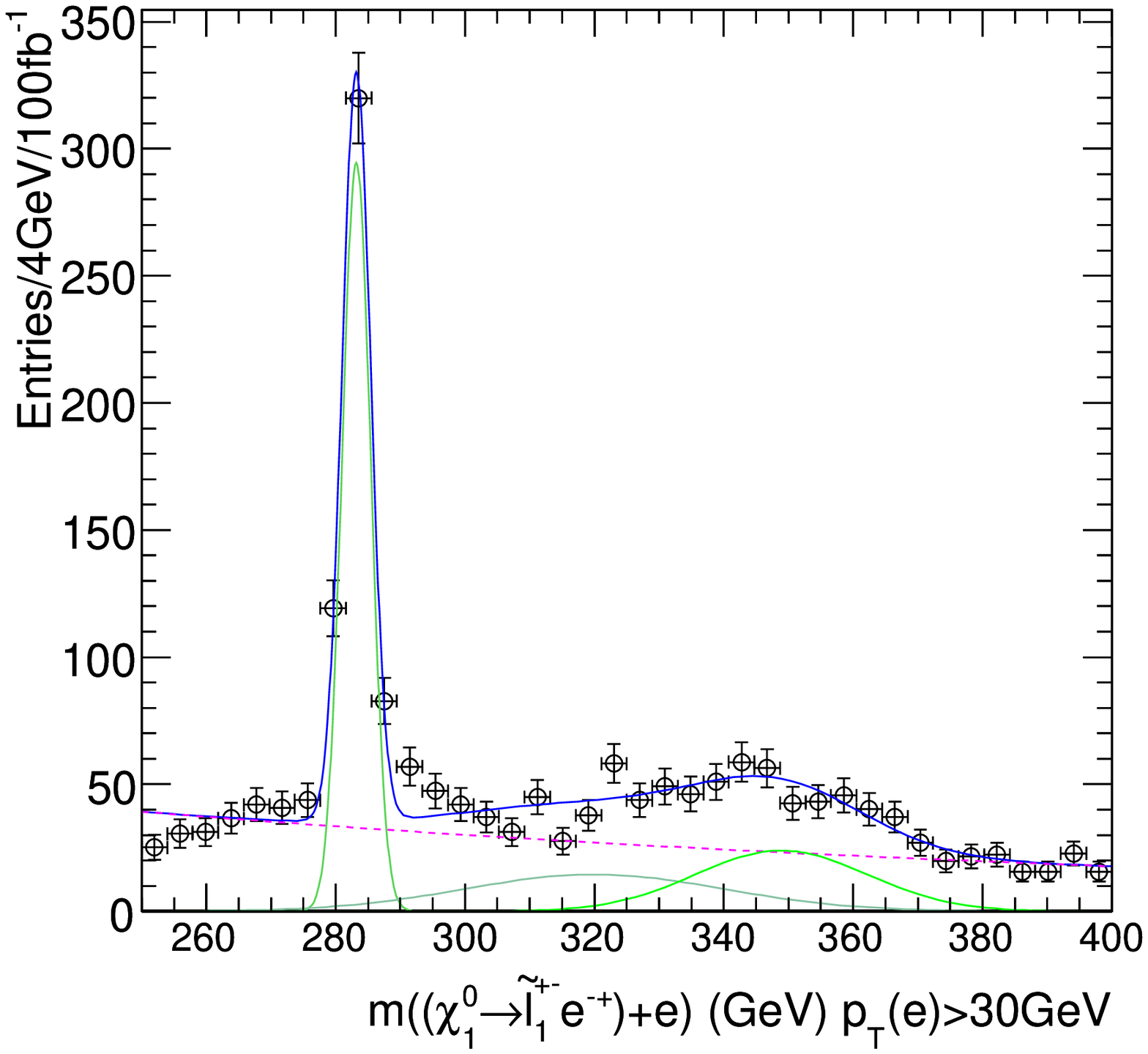} 
}
\subfigure[ $\ (\ell_1, \ell')=(\mu, \mu)$]{
\includegraphics[width=0.44\textwidth,clip]{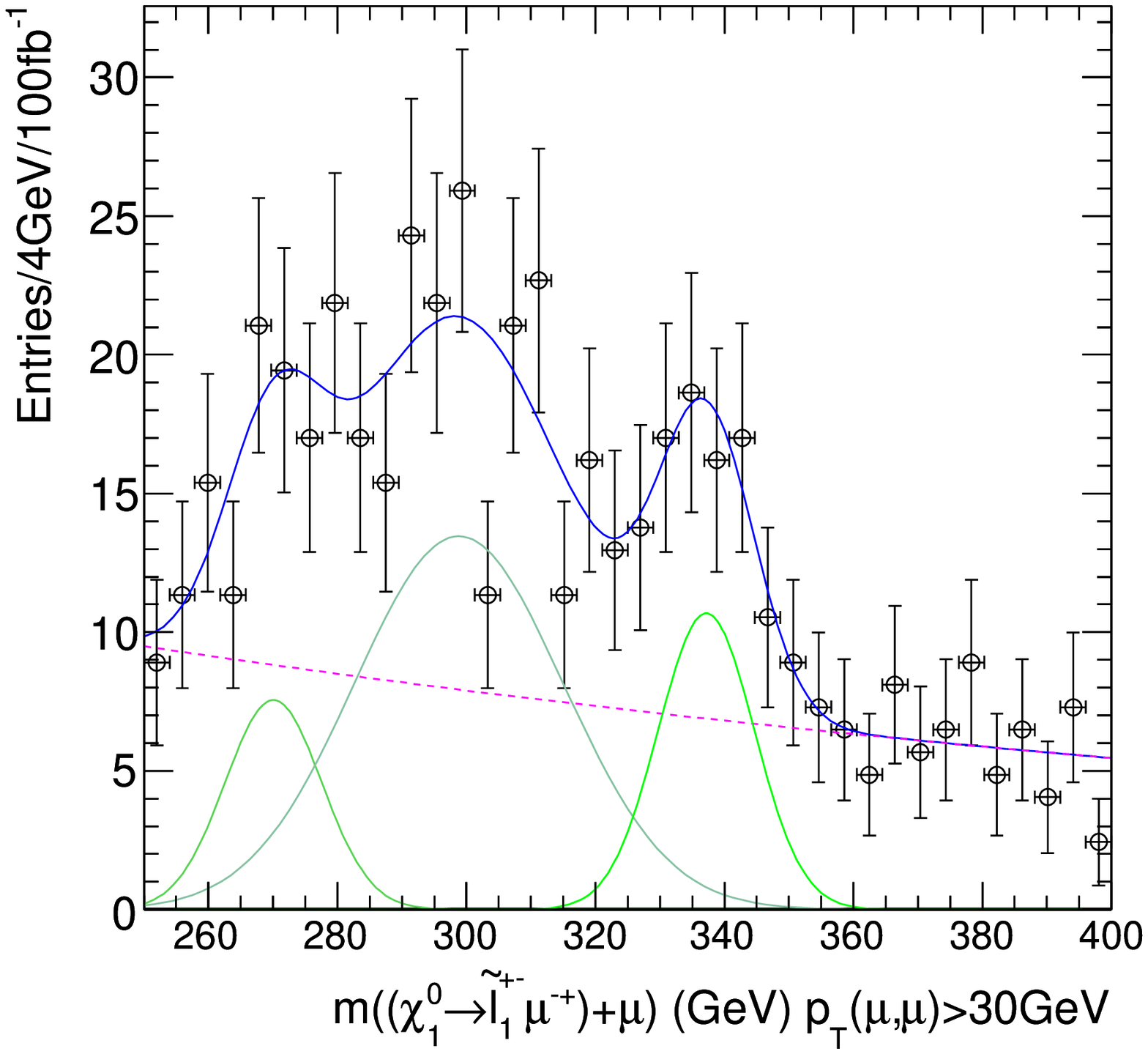} 
}
\subfigure[ $\ (\ell_1, \ell')=(\mu, e)$]{
\includegraphics[width=0.44\textwidth,clip]{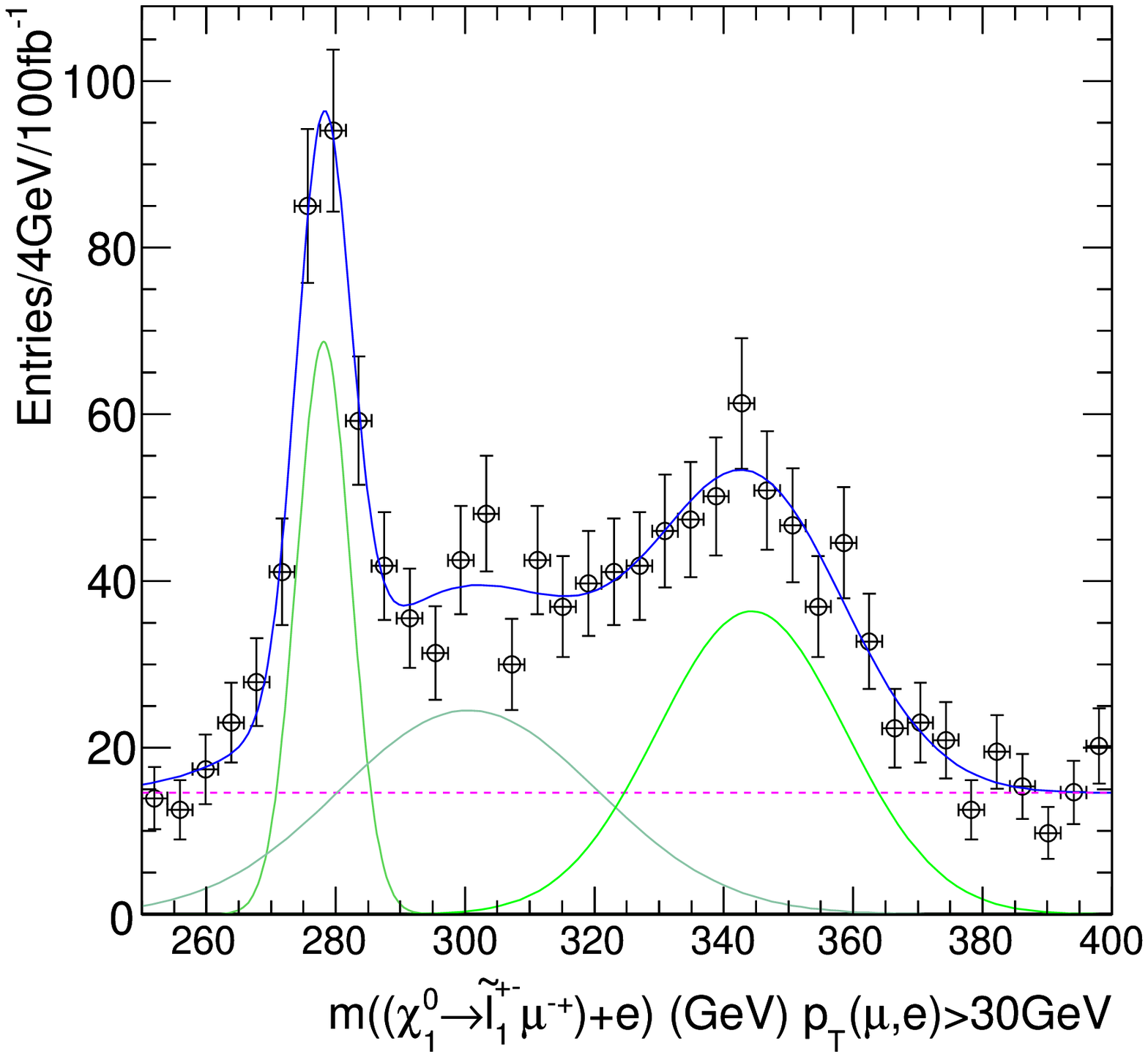}
}
\caption{\label{fig:OS_slep456} \textit{OS Slepton-Dilepton Invariant
    Mass Distributions.}  Invariant mass distributions of
    $(\slepton^{\pm}_{1}\, {\ell_1}^{\mp}) \ell'$, where the
    $\slepton^{\pm}_1 \ell_1^{\mp}$ pair reconstructs $\chi_1^0$ as
    described in the text, and $(\ell_1, \ell') =$ (a) $(e, \mu)$, (b)
    $(e, e)$, (c) $(\mu, \mu)$, and (d) $(\mu, e)$.  Both $\ell_1$ and
    $\ell'$ are required to have $p_T > 30~\gev$.  These distributions
    in the range $250~\gev < m < 400~\gev$ have been fit with Gaussian
    peaks on top of an exponentially decaying background as given by
    the green (lighter) solid and purple (lighter) dashed contours,
    respectively.  The sum of these fits is given by the blue (darker)
    solid line.}
\end{center}
\end{figure}

However, this peak starts at around 140 GeV, which, as we already
know, is the $\slepton_2$ mass. We therefore expect another peak
around 140 GeV, originating from $\slepton_2 \to \slepton_1$ decays
with one soft lepton.  In a real detector, smearing effects would then
make it hard to conclusively establish the identity of the 160 GeV
peak.  We therefore leave the question of $\slepton_3$ identification
for future work, in which tau leptons are carefully treated.

\begin{figure}[tbp]
\begin{center}
\subfigure[ $\ (\ell_1, \ell')=(e, \mu)$]{
\includegraphics[width=0.44\textwidth,clip]{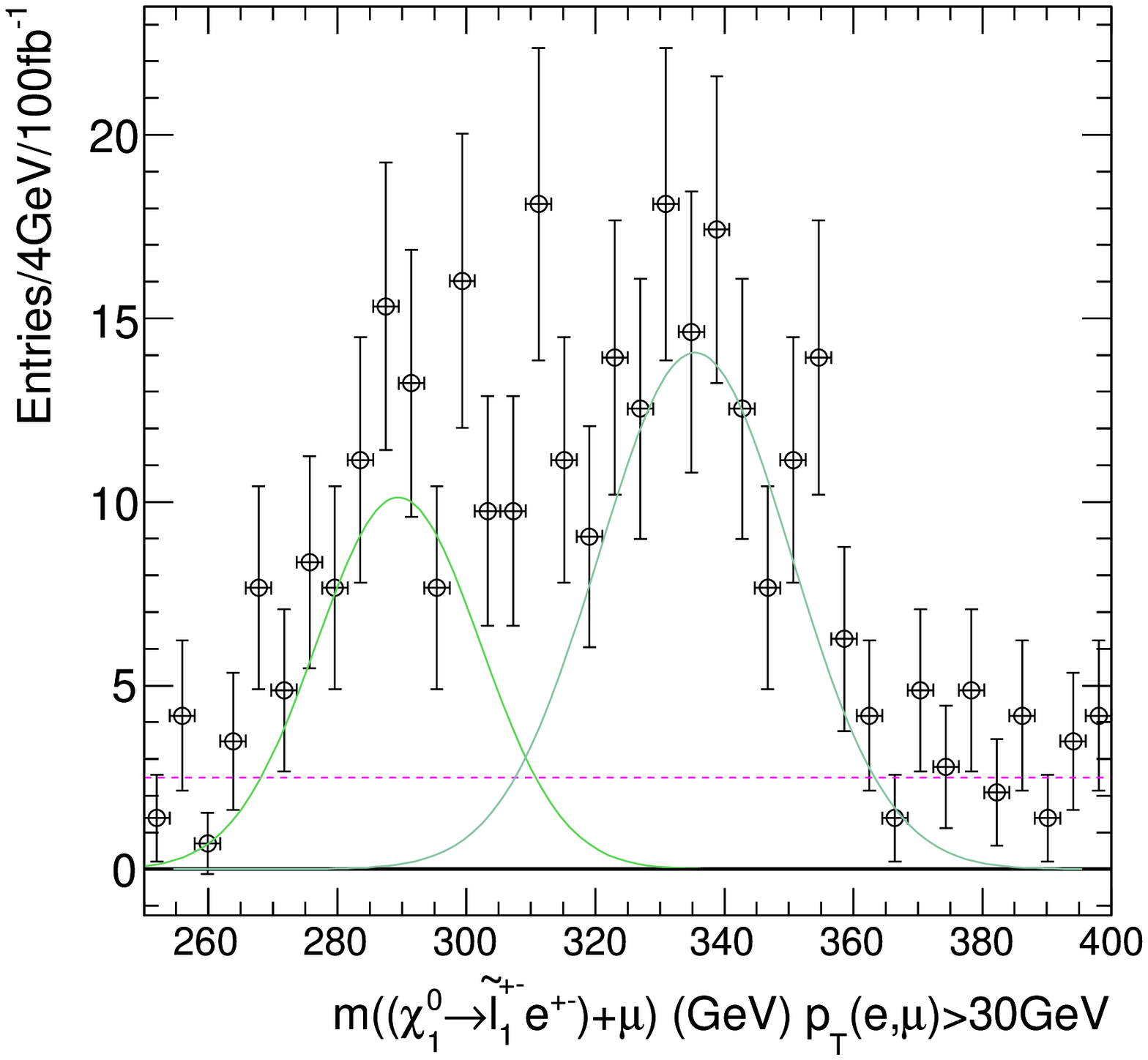} 
}
\subfigure[ $\ (\ell_1, \ell')=(e, e)$]{
\includegraphics[width=0.44\textwidth,clip]{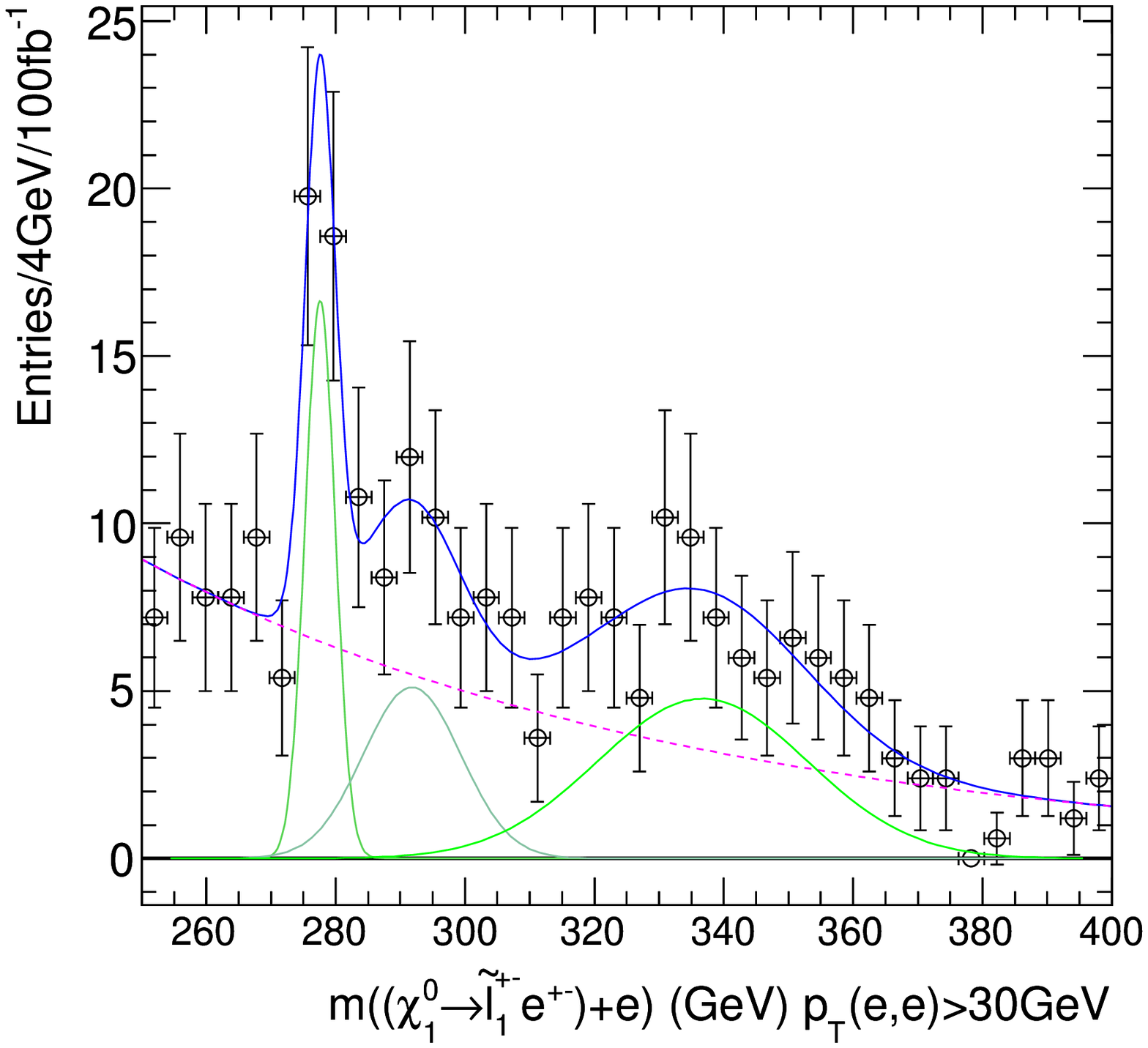}
}
\subfigure[ $\ (\ell_1, \ell')=(\mu, \mu)$]{
\includegraphics[width=0.44\textwidth,clip]{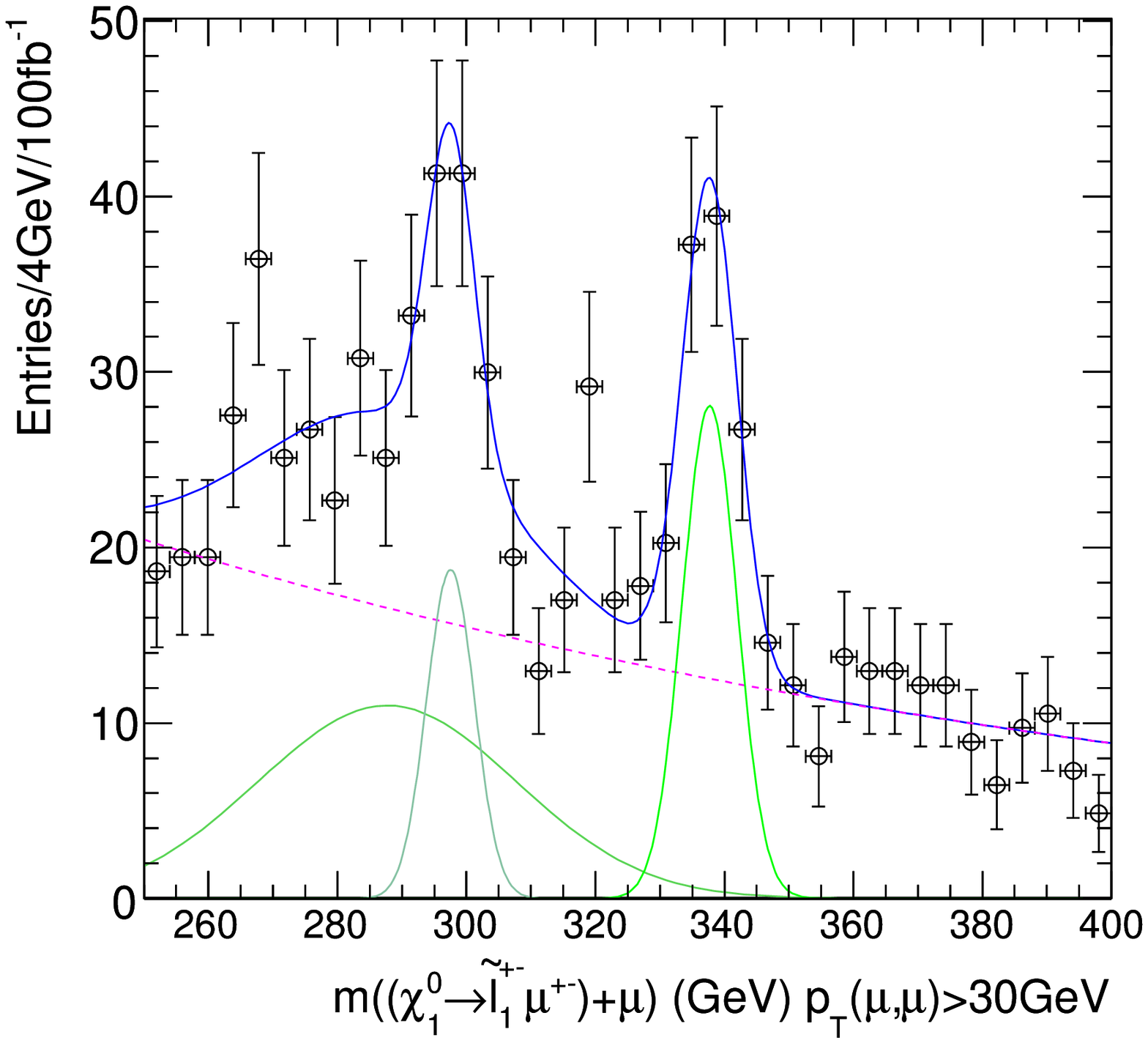}
}
\subfigure[ $\ (\ell_1, \ell')=(\mu, e)$]{
\includegraphics[width=0.44\textwidth,clip]{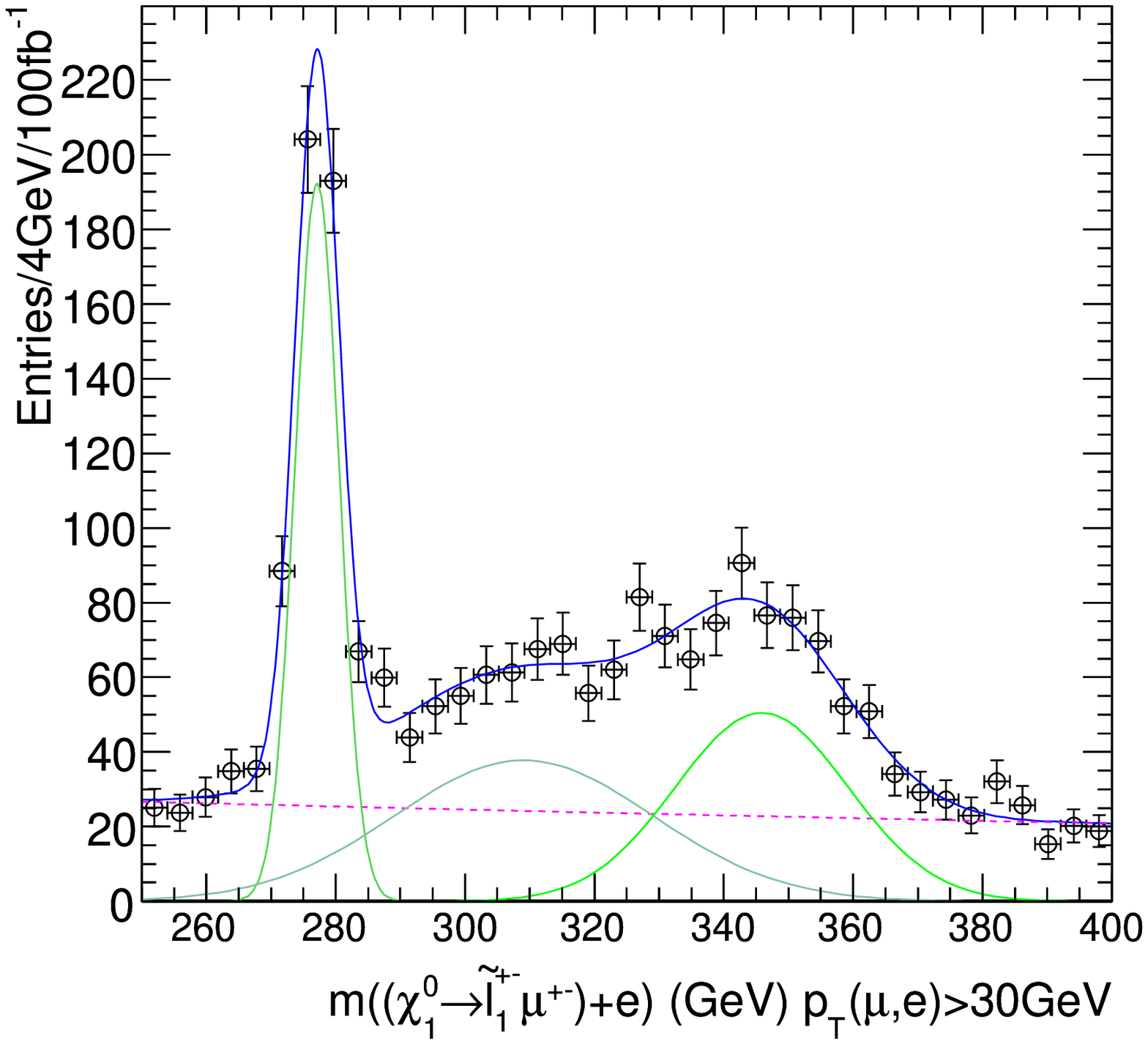}
}
\caption{\label{fig:SS_slep456} \textit{SS Slepton-Dilepton Invariant
Mass Distributions.}  As in \figref{OS_slep456}, but for SS
slepton-dilepton combinations $(\slepton^+_1\, {\ell_1}^+) {\ell'}^\pm$
and $(\slepton^-_1\, {\ell_1}^-) {\ell'}^\pm$.}
\end{center}
\end{figure}

\subsection{Reconstructing the heavy sleptons {\boldmath$\slepton_{4,5,6}$}}

Once the $\chi^0_1$ is reconstructed, we can use it as a base for
reaching higher up the decay chain. As evident in \figref{spectrum},
the heavier sleptons dominantly decay through $\slepton_{4,5,6} \to
\chi^0_1 \ell_2$.  Since in our case no heavy slepton is purely stau,
all three of the sleptons are in principle accessible even by
reconstructing only electrons and muons.

We take all $\slepton_1^\pm\, \ell_1^\mp$ combinations lying within $2
\sigma$ of the mean of the OS $\slepton_1 \ell_1$ invariant mass
distributions of \figref{OS_neutralino} and combine them with yet
another lepton $\ell' = e, \mu$ to obtain $(\slepton_1^\pm \ell_1^\mp)
\ell'$ invariant mass distributions.  Here and in the following we
enclose $\slepton_1 \ell_1$ in parentheses to indicate a
$\slepton_1$-lepton pair that reconstructs the $\chi^0_1$.  We require
that both $\ell_1$ and $\ell'$ have $p_T > 30~\gev$.  This essentially
gives us the $m_{\chi_1^0 {\ell'}}$ invariant mass distribution.  These
invariant mass distributions, with fitted peaks, are shown in
\figref{OS_slep456}.

We can also take all combinations of SS $\slepton_1$-lepton pairs with
invariant masses falling within 2$\sigma$ of the SS primary
$\slepton_1$-lepton invariant mass distribution means and combine them
with another lepton.  The resulting invariant mass distributions, with
fitted peaks, are plotted in \figref{SS_slep456}. Again we require
that both leptons have $p_T > 30~\gev$.  We know that these
combinations give a poorer reconstruction of $\slepton_{4,5,6}$ than
the OS combinations, since SS $\slepton_1 \ell_1$ pairs reconstruct not
the neutralino mass but the neutralino shifted peak, but we include
these plots for completeness.  Note that there is another reason why
we expect these plots to give an inferior reconstruction of the
heavier sleptons compared to the OS plots of \figref{OS_slep456}. The
combinatoric background is larger here, since it includes $\slepton_1$
plus two lepton combinations with charges summing to either $\pm1$ or
to $\pm3$, whereas the OS plots only include the former.

\subsection{Reconstructing the heavier neutralino {\boldmath$\chi^0_2$}}

With every step up the chain, the reconstruction errors compound,
making $\chi^0_2$ recovery harder than $\chi^0_1$. In this analysis we
do not attempt to reconstruct the heavier neutralino. Nevertheless, we
note that imposing the additional constraint that some lepton
combinations reconstruct the $\chi^0_2$ may provide additional
information and allow constraints on flavor mixings among the heavy
sleptons.

\section{Results}
\label{sec:results}

We now take our invariant mass distributions and interpret them in
terms of slepton masses and mixings.  Our best measurements of slepton
masses and mixings in this model are summarized in
\tableref{FinalResults}, and the individual entries are supported by
the detailed discussions below.

\begin{table}[tbp]
\centering
\begin{tabular}{|c|c|c|c|}
	\hline
 & True & Measured & Observation\\
	\hline
	\hline
$\slepton_1$ & 135.83 GeV & $135.9\pm0.1$ GeV & direct observation of
$\slepton_1$ with $0.6<\beta(\slepton_{1})<0.8$ (\figref{slepton1}b) \\
$\chi^{0}_{1}$ & 224.83 GeV & $225.10 \pm 0.04$ GeV & 
$\chi^0_1$ peak in the $\slepton^{\pm}_1{e}^{\mp}$ invariant mass
distribution (\figref{OS_neutralino}a) \\
$\Delta m(\slepton_{1,2})$ & 4.95 GeV & $5.06\pm 0.06$ GeV & 
$\slepton^{\pm}_1e^{\mp}$ minus $\slepton^{\pm}_1\mu^{\pm}$ peak 
positions (\figref{OS_neutralino}a, \figref{SS_neutralino}b) \\
$\slepton_{4}$ & 282.86 GeV & $283.2\pm0.1$ GeV & 
peak in $(\slepton^{\mp}_1e^{\pm})e$ invariant mass distribution (\figref{OS_slep456}b) \\
$\slepton_{5}$ & 303.41 GeV & $307\pm5$ GeV & 
peak in $(\slepton^{\mp}_1e^{\pm})\mu$ invariant mass distribution (\figref{OS_slep456}a)\\
$\slepton_{6}$ & 343.53 GeV & $343.3\pm0.5$ GeV & 
peak in $(\slepton^{\mp}_1e^{\pm})\mu$ invariant mass distribution (\figref{OS_slep456}a)\\
$|U_{2e} / U_{2\mu}|^2$ & 0.069 & $0.071\pm0.010$ & 
$N(\slepton^{\pm}_1e^{\pm})/N(\slepton^{\pm}_1\mu^{\pm})$ (\figref{SS_neutralino}) \\
	\hline
\end{tabular}
\caption{\label{table:FinalResults} \textit{Results.}  Best
measurements of slepton masses and mixings in our model.  The
measurement for $\slepton_{5}$ is given under the assumption that the
excess of events we see is indeed a mass resonance, and not a
fluctuation of the background.}
\end{table}

\subsection{The light sleptons {\boldmath$\slepton_{1,2,3}$} and the
  lightest neutralino {\boldmath$\chi^0_1$}}

As explained in \secref{slepton3} we do not attempt to measure the
$\slepton_3$ mass in this analysis, although we see hints that its
mass lies somewhat above 160 GeV.  We therefore only present here
measurements of the masses of $\slepton_{1}$ and $\slepton_{2}$.  The
lack of {\em direct} observation of a third slepton is entirely to be
expected for a spectrum with mixings such as ours, and for an analysis
not attempting to reconstruct taus.

{}From the distribution given in \figref{slepton1}, we measure
$m_{\slepton_{1}} = 135.9 \pm 0.1~\gev$. Given an integrated
luminosity of $100~\ifb$, the high number of reconstructed
$\slepton_1$ allows us to measure the $\slepton_1$ mass with small
statistical error.

We expected to be able to measure the $\slepton_{2}$ indirectly using
indirect SS and indirect OS $\chi^0_1$ decays. In
\figref{OS_neutralino}a we see the invariant mass distribution of OS
$\slepton_{1} e$ pairs. This distribution is peaked at $225.10 \pm
0.04~\gev$. We expect this peak to be a combination of pure $\chi^0_1$
from direct decays and off-reconstructed $\chi^0_1$ from indirect
decays. The OS $\slepton_{1} \mu$ distribution of
\figref{OS_neutralino}b peaks at a mass roughly $5~\gev$ lower. In
\figref{SS_neutralino} we see that both SS slepton-lepton invariant
mass distributions have a mean close to the lower of the two OS peak
positions. We therefore have a measurement consistent with
Ref.~\cite{Feng:2009yq}: nearly degenerate $\slepton_{1}$ and
$\slepton_{2}$ with the $\slepton_{1}$ having a much greater selectron
component than the $\slepton_2$ (seen from the minimal contamination
of the $\chi^0_1$ peak in the OS $\slepton_1 e$ distribution).  With
this hypothesis we can identify the position of the higher peak,
namely the OS $\slepton_1 e$ peak, as the $\chi^0_1$. This gives us a
$\chi^0_1$ mass of $225.10 \pm 0.04~\gev$.

As explained in \secref{sssection}, it is better to extract
$E_{\text{shift}}$ from the SS distributions, as these only come from
indirect decays.  {}From \figref{SS_neutralino}b, we see that the
shifted peak is at $219.39 \pm 0.06~\gev$.  We then find
$E_{\text{shift}} = 5.71 \pm 0.07~\gev$.  Using \eqref{eshift}, we
then find $\Delta m \equiv m_{\slepton_2} - m_{\slepton_1} = 5.06 \pm
0.06~\gev$,  and so $m_{\slepton_2} = 141.0 \pm 0.1~\gev$.
Note that the uncertainty in $m_{\slepton_2}$ is dominantly from the
uncertainty in $m_{\slepton_1}$.

Since the shifted neutralino peak appears in both the $\slepton_1^\pm
\mu^\pm$ and $\slepton_1^\pm e^\pm$ invariant mass distributions, we
have strong evidence that $\slepton_2$ has both smuon and selectron
components.  Similarly, since we see peaks in both the $\slepton_1^\pm
e^\mp$ and $\slepton_1^\pm \mu^\mp$ distributions, we conclude that
$\slepton_1$ has both selectron and smuon components.  It is still
unclear whether these states also have stau components. As we will see
in~\secref{lightsleptons}, we cannot rule out this possibility, but
with some mild assumptions about the neutralino couplings to
$\slepton_{1,2}$ we can show that if there are non-zero stau
components in $\slepton_{1,2}$, they must be equal to each other.

The cleanest ``flavor'' measurement we can extract is based on the SS
shifted neutralino peaks in \figref{SS_neutralino}, since these are
only sensitive to $\slepton_2$.  If we assume that the neutralino
couplings to the first two generations are the same, then dividing the
number of events $N_{SSe}$ in the $\slepton_1 e$ peak of
\figref{SS_neutralino}a by the number of events $N_{SS\mu}$ in the
$\slepton_1 \mu$ peak of \figref{SS_neutralino}b, and adjusting for
the different reconstruction efficiencies for electrons and muons, we
find the ratio of selectron to smuon components of $\slepton_2$ to be
\begin{equation}\label{12mix}
|U_{2e}/U_{2\mu}|^2= N_{SSe}/N_{SS\mu}=0.071 \pm 0.010 \ ,
\end{equation}
in good agreement with our input model.  

One could try to rule out a third light slepton with a selectron or
smuon component, based on the absence of a clear peak in our
distributions.  This would support the hypothesis that
$\slepton_{1,2}$ are selectron-smuon mixtures.  Indeed, the fact that
we only observe one shifted neutralino peak in the $\slepton_1e$ and
$\slepton_1\mu$ distributions indicates that there is no such third
slepton which is close in mass to $\slepton_1$.  Furthermore, although
we have not shown these distributions here, if one does not impose the
requirement that $\slepton_1$ and one lepton reconstruct the
neutralino peak, there is no clear peak in the $\slepton_1 \ell \ell'$
invariant mass distributions, with $\ell, \ell'=e, \mu$. However, the
peak we see near 160~GeV makes it hard to conclusively exclude such a
third slepton, although, as we explained in~\secref{slepton3}, this
peak seems consistent with a pure stau state with mass above 160~GeV.

\subsection{The heavy sleptons {\boldmath$\slepton_{4,5,6}$}}

In principle, the invariant mass distributions of $(\slepton_1 \ell)
\ell'$ combinations contain a lot of information about
$\slepton_{4,5,6}$. There is however a lot of supersymmetric
background which makes the identification of the peaks
challenging. The cleanest peaks are obtained from OS combinations
$(\slepton_1^\pm e^\mp) \ell'$, where the $\slepton_1^\pm e^\mp$ pair
reconstructs $m_{\chi_1^0}$, as shown in \figref{OS_slep456}ab.  The
OS combinations $(\slepton_1^\pm \mu^\mp) \ell'$ of
\figref{OS_slep456}cd also yield useful information, although the
peaks are not as clean. Similarly, useful but not particularly clean
peaks are found in the SS invariant mass distributions of
\figref{SS_slep456}.

When we consider the invariant mass distributions of $\chi_1^0 e^\pm$
combinations for $\chi^0_1$ reconstructed from $\slepton_1^{\mp}
e_1^{\pm}$ (\figref{OS_slep456}b), we obtain a very clear peak which
identifies $\slepton_4$, with a mean value of $283.2 \pm
0.1~\gev$. The peak near the $\slepton_6$ mass produced by considering
these combinations is questionable and contains much fewer events than
the $\slepton_4$ peak. The $\slepton_5$ peak is negligible. This
strongly suggests that $\slepton_4$ has a dominant selectron
component. The invariant mass distributions of $\chi_1^0 e^\pm$
combinations for $\chi^0_1$ reconstructed from
$\slepton_1^{\mp}\mu_1^{\pm}$ (\figref{OS_slep456}d) are not as
convincing, but the presence again of a dominant peak near
$\slepton_4$ further supports $\slepton_4$ having a strong selectron
component.

When we consider the invariant mass distributions of $\chi_1^0
\mu^\pm$ combinations (\figref{OS_slep456}a), one peak is present,
which identifies $\slepton_6$. We see an additional excess of events
at a lower mass, which could be evidence for $\slepton_5$. In
\figref{SS_slep456}c a similar excess, shifted to lower mass, supports
the hypothesis that this excess of events is more than just a
fluctuation of the background and is indeed $\slepton_5$. We have
already described a lack of evidence for selectron components in these
two mass states.  The mass of the clear heavier peak, $\slepton_6$, is
$343.3\pm0.5$~GeV. If we assume the excess at lower mass is from
$\slepton_5$, then the peak describes a mass of $307\pm5$~GeV.  These
two masses are a reasonable measure of the true slepton masses given
that the model for the background is in this instance not
ideal. We would, however, need a greater number of events to truly
believe that the excess near 305~GeV is the $\slepton_5$ and not
fluctation of the background.

Assuming that $\chi^0_2$ is predominantly gaugino, and that
$\slepton_5$ and $\slepton_6$ have identical quantum numbers, the
$\chi^0_2$ branching ratios to these two sleptons can only differ due
to phase space effects.  We know that $\slepton_{5,6}$ have negligible
selectron components, so can only be smuon-stau mixtures.  The ratio
of the number of events in the $\slepton_5$ peak to the number of
events in the $\slepton_6$ peak of \figref{OS_slep456}a therefore
gives the smuon-stau mixing in these states up to the phase-space
factor. 
Although the compounded reconstruction errors and ignorance of
systematics make exact results from this analysis suspect, we can
conclude that this mixing is ${\cal O}(1)$, if we support the excess
of events near 307~GeV being a $\slepton_5$ mass peak.


\section{Discussion}
\label{sec:discussion}

We now reflect on the measurements we have made and on the
measurements we have been unable to make. In what follows, we consider
how the lessons learned from considering this particular supersymmetry
model generalize to a wider range of supersymmetric models. We also
consider in this section what we might be able to deduce if, in
addition to these experimental observations, we make the assumption that
we are expecting to observe the six slepton states of an MSSM-type
model.

\subsection{Mass Measurements}

Out of the six sleptons, we have been able to identify in a rather
convincing way the masses of four: $\slepton_1$, $\slepton_2$,
$\slepton_4$ and $\slepton_6$. We have evidence for $\slepton_5$,
though given the statistics available it is not clear that were this
peak truly observed in an experiment it would necessarily be more than
background fluctuations. The $\slepton_2$ mass has been determined
indirectly using the neutralino shifted peak, whereas the other
sleptons are found by direct measurement.  In all five cases the
masses we obtain are in reasonable or good agreement with the true
slepton masses.  We see hints for $\slepton_3$ but cannot measure its
mass conclusively because of its dominant stau component, and because
it happens to be close in mass to $\slepton_2$.

The correct determination of as many as five of the six sleptons tells
us that for fully reconstructible supersymmetric decay chains ending
in stable charged NLSPs, working up the decay chain and reconstructing
invariant mass distributions in stages is a promising
way of measuring superpartner masses in the slepton sector. We have
seen that, even with nearly degenerate states, indirect methods do
exist which may still render these slepton masses measurable.

\subsection{Mixings}

\subsubsection{Light sleptons}\label{sec:lightsleptons}

Our best mixing measurement is of the $e-\mu$ ratio in $\slepton_2$,
shown in \eqref{12mix}.  If we assume that $\slepton_{1,2}$ are
selectron-smuon mixtures, with no stau components, we have
\begin{equation}
\left( \begin{array}{c} \slepton_1 \\ \slepton_2 \end{array} \right)
= \left( \begin{array}{cc} \cos\theta^R_{12} & \sin\theta^R_{12} \\
-\sin\theta^R_{12} & \cos\theta^R_{12} \end{array} \right)
\left( \begin{array}{c} \tilde{e}_R \\ \tilde{\mu}_R \end{array} \right)
\ .
\end{equation}
With this assumption, our measurement of \eqref{12mix} implies
\begin{equation}\label{twostate}
\sin^2\theta^R_{12}=0.066 \pm 0.009 \ .
\end{equation}

How well can we test the assumption that $\slepton_{1,2}$ have no stau
component?  Clearly a direct determination cannot be done in an
analysis that does not look at $\tau$ leptons, but it is interesting
to ask whether we can use the information we have on $\slepton_1$ and
$\slepton_2$ to argue this.  As we will now see, with some mild
assumptions about the model, we can show that $\slepton_1$ and
$\slepton_2$ can only have {\it equal} stau components, i.e., $|
U_{1\tau}|=| U_{2\tau}|$.  We cannot exclude, however,
$| U_{1\tau}| =| U_{2\tau}| \neq0$.

We begin by assuming that the neutralino coupling to the three light
sleptons is flavor-blind, an assumption that is valid not only in our
model, but generically in the types of models we are considering.  In
the mass distributions of \figsref{OS_neutralino}{SS_neutralino}, we
take the number of events in the OS electron (muon) peak to be
$N_{OSe}$ ($N_{OS\mu}$).  Then
\begin{eqnarray}
N_{SSe}&=& N\, f_{SS} \, |U_{2e}|^2\\
N_{SS\mu}&=& N\, f_{SS} \, |U_{2\mu}|^2\\
N_{OSe}&=& N \left[ f_{12} \, 
|U_{1e}|^2+(1-f_{SS})\, |U_{2e}|^2 \right] \\
N_{OS\mu}&=& N \left[ f_{12} \, 
|U_{1\mu}|^2\, +(1-f_{SS})\,| U_{2\mu}|^2 \right] \ ,
\end{eqnarray}
where $N$ is the total number of neutralino decays to $\slepton_2$,
and $f_{SS}$ is the fraction of $\slepton_2\to \slepton_1$ decays in
which the slepton charge flips sign ($\slepton_2^\pm\to
\slepton_1^\mp$).  We include the factor $f_{12} \approx 1$ to account
for the different masses of $\slepton_1$ and $\slepton_2$.  Assuming
that we already know that the neutralino is a fermion and the sleptons
are scalars, we have
\begin{equation}\label{f12}
f_{12}=\left(\frac{1-m_1^2/M^2}{1-m_2^2/M^2} \right)^2 \ , 
\end{equation}
where one power comes from phase space, and the other one from the
matrix element for the $\chi^0_1\to\slepton \ell$ decay.

It is useful to combine these four equations to obtain two
combinations that are independent of $N$ and $f_{SS}$.  One such
combination is the ratio $N_{SSe}/N_{SS\mu}$, which we already used to
extract $|U_{2e}| / |U_{2\mu}|$.  The other one is
\begin{equation}
\label{withf12}
\frac{N_e}{N_\mu}=\frac{N_{SSe}+N_{OSe}}{N_{SS\mu}+N_{OS\mu}}=
\frac{f_{12}| U_{1e}|^2+| U_{2e}|^2}
{f_{12}| U_{1\mu}|^2+| U_{2\mu}|^2} \ .
\end{equation}
For the 2-state assumption, using the value of $\sin^2\theta_{12}^R$
we found above, the theoretical prediction is $N_e/N_\mu \approx
f_{12}=1.09$, in reasonable agreement with the data,
\begin{equation}\label{data}
N_e/N_\mu= 1.16 \pm 0.02 \ .
\end{equation}
It is easy to see, however, that this agreement would persist for any
\begin{equation}
|U_{2e}| ,|U_{1\mu}| \ll |U_{1e}| \simeq |U_{2\mu}| \ ,
\end{equation}
which, for $|U_{1\tau}|, |U_{2\tau}| \not\ll |U_{1e}|, |U_{2\mu}|$,
implies (by unitarity) that $|U_{1\tau}|\simeq|U_{2\tau}|$.

\subsubsection{Heavy sleptons}

It is encouraging that, despite the $\slepton_4$ being produced high
up in our decay chains, we have good evidence that $\slepton_4$ has a
very strong selectron component and negligible smuon component from
the strong peaks in the $\chi^0_1 e$ invariant mass distributions and
the absence of a peak in the $\chi^0_1 \mu$ invariant mass
distributions. 
For $\slepton_5$ and $\slepton_6$ we know that these
have strong smuon components but negligible selectron components.
Assuming that these states are left-handed sleptons, and that the
$\chi^0_2$ is a gaugino, we can deduce that the smuon-stau mixing in
these states is of ${\cal O}(1)$.  Assuming further that $\slepton_4$
is a left-handed slepton, we can also conclude that it is
predominantly a selectron.

\section{Conclusions}

The understanding of slepton masses and mixings that we have arrived
at, based on experimental evidence, is summarized in
\tableref{FinalResults}.  The results are very close to the assumed
underlying input model.  We have taken a ``data-driven'' approach, in
which we try to infer from the data the masses and flavor compositions
of new particles, rather than fitting the parameters of a particular
model to the data.  The methods we described are specific to models
with metastable charged particles and, therefore, no missing energy.

The most obvious way to identify the sleptons in our model (which
we initially tried) would be to look for peaks in the NLSP plus two
leptons invariant-mass distributions.  This does not work very well,
however, either because some of the relevant leptons are too soft or
because of the large supersymmetric combinatoric backgrounds.
Instead, we showed that the best mass measurements in these scenarios
can be obtained by reconstructing the spectrum in stages, starting
from the bottom up. For example, rather than discovering a heavy
slepton by searching for a peak in the invariant mass distribution of
the NLSP plus two leptons, we first identify the lightest neutralino
as a peak in the NLSP plus lepton distribution, and then combine this
neutralino with one additional lepton to obtain the heavier slepton
peak.  We also demonstrated that nearly-degenerate particles can be
resolved indirectly, using the shifted peak of their mother particle.

We then outlined methods for measuring the slepton flavor
compositions, essentially by counting experiments.  We were able to
qualitatively infer the flavor makeup of the sleptons, and to
quantitatively measure one mixing.

The charged NLSP is a crucial ingredient in our analysis, yet the
model we chose also has some unfavorable features:
\begin{enumerate}
\item The two lightest sleptons are almost degenerate.
\item The $e-\mu$ mixing in these sleptons is small.
\item One slepton, $\slepton_3$, is an almost pure stau state, and is
quite close in mass to $\slepton_{1,2}$, so that the peak it gives in
the NLSP plus $e/\mu$ distributions is very close to the $\slepton_2$
peak.
\end{enumerate}
We have also not used any information from tau leptons.  Given these
difficult features and the pessimistic assumption about taus, the
amount of information we were able to extract is encouraging.

The analysis we presented is certainly preliminary.  If our toy model
were real data, one would want to perform detailed measurements to
confirm that the observed events were supersymmetric, establish the
gaugino identities of the neutralinos, and explore the squark and
gluino sector.  One could then fit the flavor parameters of the model
to the data, and flesh out the details of the picture we outlined
here.

We conclude that, under favorable but not outrageously optimistic
circumstances (unless we think of discovering supersymmetry as
outrageously optimistic), it is possible that measurements made at the
Large Hadron Collider will allow us to determine a great deal about
the slepton spectrum and flavor decomposition, and constrain, and
possibly clarify, the flavor structures of both supersymmetry and the
standard model.

\section*{Acknowledgments}

We thank James Frost and Are Raklev for technical assistance and Kfir
Blum for discussions.  YS thanks the Aspen Center for Physics, where
part of this work was completed.  The research of JLF, YN, YS and IG
was supported in part by the United States-Israel Binational Science
Foundation (BSF) under Grant No.~2006071.  The work of JLF, DS, and FY
was supported in part by NSF Grant No.~PHY--0653656.  STF and CGL
acknowledge the support of the United Kingdom's Science and Technology
Facilities Council, Peterhouse and the University of Cambridge.  The
work of IG and YS was supported in part by the Israel Science
Foundation (ISF) under Grant No.~1155/07.  The work of YN is supported
by the Israel Science Foundation (ISF) under Grant No.~377/07, the
German-Israeli Foundation for scientific research and development
(GIF), and the Minerva Foundation.

\appendix

\section{Model Details}
\label{sec:details}

\subsection{Input Parameters}

As discussed in \secref{model}, the model used for this analysis is a
hybrid model in which slepton masses receive contributions from both
gauge- and gravity-mediated supersymmetry breaking.  Gauge-mediation
provides the primary, flavor-universal contributions to soft masses,
and gravity-mediation supplements these with flavor-violating
contributions.  The slepton soft mass matrices then have the form
\begin{eqnarray}
M_{\tilde{\nu}}^2 &=& m_{\slepton}^2 {\mathbf 1} + x m_{\slepton}^2
X_L \\
M_{\slepton_L}^2 &=& m_{\slepton}^2 {\mathbf 1} + m_E m_E^{\dagger}
+ x m_{\slepton}^2 X_L \\ 
M_{\slepton_R}^2 &=& m_{\tilde{R}}^2 {\mathbf 1} + m_E^{\dagger} m_E
+ x m_{\slepton}^2 X_R \ ,
\end{eqnarray}
where $m_{\slepton}^2$ and $m_{\tilde{R}}^2$ are the gauge-mediated,
flavor-conserving contributions to the left- and right-handed
sleptons, $m_E$ is the lepton mass matrix, and $x$ is the ratio
between gravity- and gauge-mediated contributions.  The symmetric
matrices $X_L$ and $X_R$ parameterize the gravity-mediated
flavor-violating effects for left- and right-handed sleptons, with the
flavor structure determined by the Froggatt-Nielsen
mechanism~\cite{Froggatt:1978nt}.

For the model analyzed here, the flavor-conserving gauge-mediated
contributions are specified by the standard GMSB parameters
\begin{equation}
\nmess = 5 , \ \mmess = 4.6 \cdot 10^6~\gev, \ \Lambda = 3.4 \cdot
10^4~\gev , \ C_{\text{grav}} = 1 , \ \tan \beta = 10 , \ \mu > 0 \ .
\end{equation}
The gravity-mediated, flavor-violating contributions are governed by a
U(1) $\times$ U(1) horizontal symmetry under which the lepton
superfields have the charges
\begin{equation}
L_1(2,0), \ L_2(0,2), \ L_3(0,2); \quad \overline{E}_1(2,1), \
\overline{E}_2(2,-1), \ \overline{E}_3(0,-1) \ .
\end{equation}
The resulting flavor-violating masses are generated using appropriate
powers of the Froggatt-Nielsen expansion parameter $\lambda$ with
${\cal O}(1)$ coefficients for each term in the matrix.  The resulting
flavor-violating contributions for the sleptons are
\begin{eqnarray}
X_L & = & \left(
\begin{array}{rrr}
-3.4989 \lambda^0 & \ -1.2001 \lambda^4 & \ 0.4059 \lambda^4 \\
-1.2001 \lambda^4 & -1.2705 \lambda^0 & 1.1746 \lambda^0 \\
0.4059 \lambda^4 & 1.1746 \lambda^0 & 1.4293 \lambda^0
\end{array} \right) \\
X_R & = & \left(
\begin{array}{rrr}
-1.0368 \lambda^0 & 0.9976 \lambda^2 & \ -0.06188 \lambda^4 \\
0.9976 \lambda^2 & \ -0.8616 \lambda^0 & 0.2204 \lambda^2 \\
-0.06188 \lambda^4 & 0.2204 \lambda^2 & 0.6544 \lambda^0
\end{array} \right) \ ,
\end{eqnarray}
and the lepton mass matrix is given by

\begin{equation}
m_E = \langle \phi_d \rangle \lambda \left(
\begin{array}{rrr}
-0.3838 \lambda^4 & 0 & 0 \\
0.8706 \lambda^4 & \ -1.8682 \lambda^2 & \ -1.5408 \lambda^0 \\
1.0450 \lambda^4 & 0.3574 \lambda^2 & 1.8554 \lambda^0
\end{array} \right) \ .
\end{equation}
We ignore the neutrino mass matrix for our purposes, as neutrino
flavor is unobservable in colliders and thus unimportant for this
study.  We take $x = 0.1$ and $\lambda = 0.2$.

\subsection{Spectrum}

Given these input parameters, the full supersymmetric spectrum and
flavor-general decay branching ratios are generated by the program
\textsc{Spice}~\cite{Engelhard:2009br}.  The \textsc{Spice} input file
used is

\noindent {\tt
\\
\indent gmsb 5 4.6e6 3.4e4 1.0 10 1 \\
\indent x 0.1 \\
\indent lambda 0.2 \\
\indent nCharges 2 \\
\indent L1 2 0 \\
\indent L2 0 2 \\
\indent L3 0 2 \\
\indent E1 2 1 \\
\indent E2 2 -1 \\
\indent E3 0 -1 \\
\indent Lep -0.3838 0 0 0.8706 -1.8682 -1.5408
1.0450 0.3574 1.8554 \\
\indent XL -3.4989 -1.2001 0.4059 -1.2001 -1.2705 1.1746 0.4059
1.1746 1.4293 \\
\indent XR -1.0368 0.9976 -0.06188 0.9976 -0.8616 0.2204 -0.06188
0.2204 0.6544 }
\\

\begin{center}
\begin{table}[tb]
\begin{tabular}{|c||c||r|r|r|r|r|r|}
\hline
& Mass (GeV) & $\tilde{e}_R \quad$ & $\tilde{\mu}_R \quad$ &
$\tilde{\tau}_R \quad$ & $\tilde{e}_L \quad$ & $\tilde{\mu}_L \quad$ &
$\tilde{\tau}_L \quad$ \\
\hline
$\slepton_6$ & 343.53 & $0$ & $-0.0043$ & $-0.0445$ & $-0.0005$
& $0.8789$ & $0.4749$ \\
$\slepton_5$ & 303.41 & $0$ & $0.0028$ & $-0.1154$ & $-0.0009$ &
$-0.4767$ & $0.8715$ \\
$\slepton_4$ & 282.86 & $0$ & $0$ & $-0.0002$ & $1.0000$ &
$0$ & $0.0010$ \\
$\slepton_3$ & 168.30 & $-0.0012$ & $0.0238$ & $0.9920$ & $0$ &
$-0.0159$ & $0.1226$ \\
$\slepton_2$ & 140.78 & $-0.2551$ & $0.9666$ & $-0.0231$ & $0$ &
$0.0053$ & $-0.0032$ \\
$\slepton_1$ & 135.83 & $ 0.9669$ & $0.2551$ & $-0.0048$ & $0$ &
$0.0014$ & $-0.0007$ \\
\hline
\end{tabular}
\caption{\textit{Physical Slepton Masses and Flavor Composition.}
Elements of the mixing matrix with values less than $10^{-4}$ have
been set to zero.}
\label{table:sleptons}
\end{table}
\end{center}

\begin{center}
\begin{table}[tb]
\begin{tabular}{|c||c||r|r|r|}
\hline
& Mass (GeV) & $\tilde{\nu}_e \quad$ & $\tilde{\nu}_\mu \quad$ &
$\tilde{\nu}_\tau \quad$ \\
\hline
$\tilde{\nu}_3$ & 334.44 & $-0.0005$ & $0.8695$ & $0.4939$ \\ 
$\tilde{\nu}_2$ & 293.82 & $-0.0009$ & $-0.4939$ & $0.8695$ \\
$\tilde{\nu}_1$ & 271.37 & $1.0000$ & $0$ & $0.0010$ \\
\hline
\end{tabular}
\caption{\textit{Physical Sneutrino Masses and Flavor Composition.}
Elements of the mixing matrix with values less than $10^{-4}$ have
been set to zero.}
\label{table:sneutrinos}
\end{table}
\end{center}

\begin{center}
\begin{table}[tbh]
\begin{tabular}{|c|r||c|r|}
\hline 
& Mass (GeV) & & Mass(GeV) \\
\hline
$\tilde{d}_L$ & 1093.69 & $h^0$ & 111.50 \\
$\tilde{u}_L$ & 1086.02 & $H^0$ & 562.02 \\
$\tilde{s}_L$ & 1093.58 & $A^0$ & 561.68 \\
$\tilde{c}_L$ & 1086.02 & $H^+$ & 567.61 \\
$\tilde{b}_1$ & 1039.56 & $\chi_1^0$ & 224.83 \\
$\tilde{t}_1$ &  944.91 & $\chi_2^0$ & 398.87 \\
$\tilde{d}_R$ & 1050.83 & $\chi_3^0$ & 470.73 \\
$\tilde{u}_R$ & 1053.09 & $\chi_4^0$ & 521.28 \\
$\tilde{s}_R$ & 1050.83 & $\chi_1^+$ & 399.53 \\
$\tilde{c}_R$ & 1053.09 & $\chi_2^+$ & 522.02 \\
$\tilde{b}_2$ & 1050.92 & $\tilde{g}$ & 1225.74 \\
$\tilde{t}_2$ & 1061.48 & $\tilde{G}$ & $\sim 40$ \\
\hline
\end{tabular}
\caption{\textit{Particle Spectrum.}  Physical masses of the squarks,
gluino, Higgs bosons, charginos, neutralinos, and gravitino.}
\label{table:spectrum}
\end{table}
\end{center}

The masses of the six slepton mass eigenstates, as well as the
rotation matrix relating the slepton mass and gauge eigenstates, are
presented in \tableref{sleptons}.  These are discussed in
\secref{model}, with a graphical representation given in
\figref{spectrum}.  The sneutrino masses and flavor compositions are
given in \tableref{sneutrinos}.

The masses of the remaining MSSM particles are given in
\tableref{spectrum}.  The SM-like Higgs boson is below the current
bounds, but it and its precise mass play no role in the signals
studied here.  The gravitino mass is not precisely specified.  The
dominant contribution to its mass is from gravity-mediation, and so we
expect $m_{\tilde{G}} \sim \sqrt{x} m_{\slepton}$.  This guarantees
that decays to the gravitino take place outside collider detectors and
the signals we study are insensitive to the precise value of
$m_{\tilde{G}}$.  The general mass ordering of the lightest states is
therefore
\begin{equation}
m_{\tilde{G}} \ll m_{\slepton_{1,2,3}} < m_{{\chi}_1^0} <
m_{\slepton_{4,5,6}} < m_{{\chi}_2^0}  \ .
\end{equation}
The decays from the lightest neutralino to the light sleptons, and
from the light sleptons to other light sleptons, are also shown in
\figref{spectrum}.

\end{document}